\documentclass[aps,prb,longbibliography,groupedaddress,twocolumn,superscriptaddress,10pt]{revtex4-2}
\usepackage{graphicx}
\usepackage{mathtools}
\usepackage{amsmath}
\usepackage{amssymb}
\usepackage{bbm}
\usepackage{bm}
\usepackage{cancel}
\usepackage{xspace}
\usepackage{braket}
\usepackage{xcolor}
\usepackage{siunitx}
\usepackage{calrsfs}
\usepackage{comment}
\usepackage{placeins}
\usepackage[export]{adjustbox}
\usepackage{wrapfig}
\usepackage{bbold}
\usepackage[colorlinks=true]{hyperref} 

\newcommand{\ee}{\mathrm{e}}  
\DeclareMathOperator*{\ii}{i} 
\newcommand*\dd{\mathop{}\!\mathrm{d}}

\renewcommand{\vec}[1]{\bm{#1}} 
\newcommand{\mat}[1]{\bm{#1}} 
\newcommand{\kel}[1]{\underline{#1}} 

\definecolor{hblue}{RGB}{0,80,255}
\definecolor{hred}{RGB}{255,80,0}

\begin{document}

\title{Impact of disorder and phonons on the Hubbard bands of Mott insulators in strong electric fields}
	
\author{Tommaso Maria Mazzocchi}
\email[]{mazzocchi@tugraz.at}
\affiliation{Institute of Theoretical and Computational Physics, Graz University of Technology, 8010 Graz, Austria}
\author{Daniel Werner}
\affiliation{Institute of Theoretical and Computational Physics, Graz University of Technology, 8010 Graz, Austria}
\author{Enrico Arrigoni}
\email[]{arrigoni@tugraz.at}
\affiliation{Institute of Theoretical and Computational Physics, Graz University of Technology, 8010 Graz, Austria}

\date{\today}
	
\begin{abstract}
We characterize the current-carrying non-equilibrium steady-state (NESS) in a single-band Hubbard model confronted with a static electric field in the presence of quenched disorder. Beyond linear response regime, the electric field amplitude must be such to compensate for at least half of the band gap in order to have a non-negligible stationary current. As disorder is not expected to dissipate the extra energy injected by the field, optical phonons assisted by a fermionic heat bath serve as dissipation channels for the current-induced Joule heat generated by the accelerated electrons. The NESS of the system is addressed by means of the dynamical mean-field theory using the so-called auxiliary master equation approach as impurity solver. Disorder effects are treated locally via the coherent potential approximation (CPA) and the self-consistent Born (SCB) approach. In the regime in which the two schemes yield similar results, we employ the SCB as it is computationally cheaper than the CPA. We show that, in a purely electronic setup, the disorder-induced dephasing cannot contribute states within the gap but only smear out the edges of the Hubbard bands. When phonons are taken into account, the different nature of disorder-induced dephasing and phonon-related dissipation becomes clear. We show that although both disorder and electron-phonon interaction enhance the current at off-resonant fields, disorder effects play a marginal role since they cannot provide in-gap states which are instead brought about by phonons and represent the privileged relaxation pathway for excited electrons.
\end{abstract}

	
\maketitle

\section{Introduction}\label{sec:intro}

Among the plethora of condensed matter phenomena, the insulator-to-metal transition (IMT) has been widely studied from both the experimental~\cite{st.ca.13,ja.tr.15} and the theoretical point of view~\cite{le.ra.85,be.ki.94}. To accomplish a deeper understanding of the microscopic processes underlying this phenomenon, progressively more complex models have been studied in the last few decades. Some of the most important ones include {\em effective models}~\cite{st.ca.13,ja.tr.15,li.ar.17}, which can describe the experimental results. However, for a comprehensive explanation of the heat dissipation more realistic mechanisms such as the interaction between the hot electrons and lattice vibrations have to be taken into account. Early attempts to come up with a coherent theoretical framework aimed at understanding the role played by a fermion bath in the context of a dissipative system driven out of equilibrium~\cite{ts.ok.09}. Fermion baths were later successfully employed to reach a {\em non-trivial} non-equilibrium steady-state (NESS)~\cite{aron.12,am.we.12,ar.ko.12,li.ar.15,li.ar.17,ha.li.18,mu.we.18} in Mott-insulating systems. More recent years have witnessed a growing interest in the role of the electron-phonon (e-ph) interaction~\cite{mu.we.15,mu.ts.17,pi.gr.23,pi.gr.23.ss,ma.ga.22,ga.ma.22,ma.we.23,ha.ar.23} in correlated systems. 
It is beyond a doubt that a satisfactory description of Joule heat dissipation due to lattice vibrations is necessary in order to clarify the nature of the IMT, as it is still debated whether the latter is due to thermal~\cite{li.ar.15,ha.li.18,di.ha.23} or quantum processes~\cite{ha.ar.23}. As of today, the most well-established nonperturbative method to deal with correlated systems is the so-called dynamical mean-field theory (DMFT)~\cite{me.vo.89,ge.ko.92,ge.ko.96,ko.sa.06,fr.tu.06,ao.ts.14} which hold under both equilibrium and non-equilibrium conditions and has been employed in most of the aforementioned studies.

An additional step toward a comprehensive description of the IMT in real materials consists in the inclusion of disorder, the theoretical framework of which has been established in~\cite{le.ra.85,zu.we.90,goni.92} after the pioneering work of Anderson~\cite{ande.58}. The latter initiated the study of correlated metals in equilibrium starting from the analysis of single-particle {\em localization} in disordered systems. However, upon inclusion of electronic correlation it is hard to tell the IMT due to purely electron-electron interaction~\cite{mott.49,im.fu.98} from the disorder-induced lozalization~\cite{ande.58,ev.me.08,te.zh.18}. In equilibrium, the possibility of having a non-zero critical temperature for the IMT in systems with interacting electrons in static random potentials has been recently investigated, e.g., in Refs.~\cite{ba.al.06,og.hu.07}. More importantly, in Ref.~\cite{og.hu.07} mention is already made of the need for a heat bath to have conduction in insulating systems given their inability to self-equilibrate, i.e. self-thermalize. It is then evident that an explanation of the Joule heat dissipation within correlated systems acquires a very practical importance other than being a fundamental question on its own.

However, even within the DMFT the study of correlated {\em disordered} systems is as yet quite challenging. The reason lies in the large number of inequivalent configurations to be dealt with, which, in turn, increases the complexity of the problem at hand.
In the context of noninteracting systems the DMFT historically originates from the so-called coherent potential approximation (CPA)~\cite{sove.67,ve.ki.68,el.kr.74} which was originally introduced to describe a coherent medium embedded in an effective environment to be self-consistently determined by requiring the averaged onsite scattering to vanish. Indeed, by means of the functional integral formulation, it has been shown that the CPA is a special case of DMFT for disordered systems~\cite{jani.89,ja.vo.92}, thus allowing the combination of the two approaches to treat both electronic correlation and disorder on the same footing. Recent works~\cite{do.te.22,ya.we.23u} have focused on the application of a combined $\text{DMFT}$ {\em plus} $\text{CPA}$ scheme to transport problems within correlated systems subject to a quench of the electron-electron interaction in a time-resolved fashion.

In this work we merge the scheme presented in~\cite{do.te.22,ya.we.23u} with our auxiliary master equation approach~\cite{ar.kn.13,do.nu.14,do.ga.15,ti.do.15,we.lo.23} (AMEA) impurity solver which allows to access the NESS of the system skipping its time evolution. The NESS is addressed by means of the DMFT and its non-equilibrium Floquet extension~\cite{jo.fr.08,ts.ok.08,so.do.18,ma.ga.22,ga.ma.22,ma.we.23}. Our main goal is the characterization of a {\em disordered} single-band Hubbard model in terms of its conducting properties with focus on the interplay between fermionic and phononic degrees of freedom. The optical phonons employed in this work are included in a {\em perturbative} fashion~\cite{ma.ga.22,ma.we.23} by means of the Migdal approximation~\cite{mu.we.15,mu.ts.17} and, together with the fermionic bath, contribute the dissipation channels used by the system to get rid of the energy injected by the field. To create states around the Fermi level, i.e. to reach the metallic phase, the electric field has to compensate for at least half of the band gap~\cite{ma.ga.22}. As a matter of fact such field amplitudes would be too strong~\cite{ec.we.13,mu.we.18,ap.st.12,ma.ga.22} to be used in experiments with a regular dc-field without damaging the material. On the other hand, as argued in Ref.~\cite{mu.we.18}, such strong electric fields can be achieved by THz field pulses which can be seen as quasi-static in comparison to the typical time scales of the system. We benchmark the CPA~\cite{do.te.22,ya.we.23u} scheme against the so-called self-consistent Born (SCB) approximation~\cite{mu.ec.18}, which we will employ throughout the whole Manuscript as it is computationally cheaper than the former.

We show that the steady-state current monotonically drops as the strength of the coupling to the fermionic bath extrapolates to zero~\cite{ma.ga.22,ma.we.23} with and without disorder, for applied fields compensating half of or the whole band gap. This confirms the expectation that elastic processes due to scattering with disorder are
not enough to sustain a steady-state current, were it not for the fermionic bath providing an energy dissipation~\cite{ma.ga.22}.
However, the disorder-induced dephasing enhances the current at off-resonant fields by producing a slight leak of the edges of the Hubbard bands into the band gap.

Finally, as far as the conducting properties are concerned we find that the effects of disorder and e-ph scattering are comparable in that they both enhance the current at off-resonant fields. On the other hand, at resonance the current is essentially not affected by the introduction of phonons and disorder. 
However, while phonons can absorb energy from the hot electrons of the lattice by providing in-gap states to {\em bridge} the band gap, disorder can only contribute dephasing effects which are negligible when the e-ph coupling is sufficiently strong.

The Manuscript is organized as follows: In Sec.~\ref{sec:MO_HA} we introduce the model of interest, while in Sec.~\ref{sec:method} we discuss the implementation of disorder within the DMFT approach. Results are presented in Sec.~\ref{sec:General_Results} while Sec.~\ref{sec:conclusions} is left for conclusions and final comments.
	
\section{Model Hamiltonian}\label{sec:MO_HA}
 
The model we are considering (see also~\cite{ec.we.13,mu.we.18,mu.ts.17}) consists in the setup described in~\cite{ma.ga.22,ma.we.23}, namely the single-band Hubbard model subject to a constant electric field {\em plus} disorder. Its Hamiltonian reads
\begin{equation}\label{eq:MicroHamiltonian}
\hat{H}(t) = \hat{H}_{\text{\tiny U}}(t) + \hat{H}_{\text{bath}} + \hat{H}_{\text{e-ph}} + \hat{H}_{\text{ph}} + \hat{H}_{\text{ph},\text{ohm}}.
\end{equation}
The Hamiltonian $\hat{H}_{\text{\tiny U}}(t)$ in presence of quenched disorder~\cite{do.te.22} and an external electric field is given by
\begin{align}\label{eq:Hubbard_ham}
\begin{split}
\hat{H}_{\text{\tiny U}}(t) & = \sum_{i\sigma} \left( \varepsilon_{\text{c}} + V_{i} \right) \hat{n}^{f}_{i\sigma} + U \sum_{i} \hat{n}^{f}_{i\uparrow} \hat{n}^{f}_{i\downarrow} \\
& -\sum_{(i,j)}\sum_{\sigma} \underbrace{t_{\text{c}} \ \ee^{-\ii \frac{Q}{\hbar} \left( \vec{r}_j - \vec{r}_i \right) \cdot \vec{A}(t)}}_{= \ t_{ij}(t)} \hat{f}^{\dagger}_{i\sigma} \hat{f}_{j\sigma},
\end{split}
\end{align}
where $\hat{f}^{\dagger}_{i\sigma}$ ($\hat{f}_{i\sigma}$) is the creation (annihilation) operator of an electron of spin $\sigma= \{ \uparrow,\downarrow \}$ at the $i$-th lattice site and $\hat{n}^{f}_{i\sigma}\equiv \hat{f}^{\dagger}_{i\sigma} \hat{f}_{i\sigma}$ the corresponding density operator. Sums over nearest neighbor sites are denoted by $(i,j)$ and the electrons' {\em onsite energy} is chosen as $\varepsilon_{\text{c}} \equiv -U/2$ such to fulfill particle-hole symmetry. 

The electric field is introduced in the temporal gauge {\em via} the Peierls substitution~\cite{peie.33,so.do.18,mu.we.18,ma.ga.22,ma.we.23} leading to a time-dependent hopping. $t_{\text{c}}$ is the {\em bare} hopping amplitude, $\vec{A}$(t) the homogeneous vector potential, $Q$ the electron charge and $\hbar$ Planck's constant. 
In the model considered~\cite{ts.ok.08,mu.we.18,ma.ga.22,ma.we.23} the electric field $\vec{F}$ lies along the lattice body diagonal $\vec{e}_{0}=(1,1,\ldots,1)$ and is given by $\vec{F}= -\partial_{t}\vec{A}(t)$ with $\vec{A}(t)=-F t \vec{e}_{0}$. Also, we define the Bloch frequency  $\Omega \equiv -FQa/\hbar$ with $a$ the lattice spacing and $F\equiv |\vec{F}|$. 

The microscopic form of the heat bath hamiltonian $\hat{H}_{\text{bath}}$ in Eq.~\eqref{eq:MicroHamiltonian} consists of a collection of noninteracting fermionic degrees of freedom (see, e.g. Refs.~\cite{aron.12,ne.ar.15,mu.we.18}) which serve as a thermostat for the accelerated electrons in the lattice. Further details about the effective contribution of $\hat{H}_{\text{bath}}$ will be provided in Sec.~\ref{sec:Dyson-eq}, see Eq.~\eqref{eq:WBL_bathGF}.

Here we consider a $d$-dimensional lattice~\footnote{The Authors are aware that this choice is quite special as it prevents the investigation of the so-called {\em dimensional crossover} occurring at the IMT, see e.g. Ref.~\cite{aron.12}. A possible setup to investigate this effect would be a two-dimensional {\em Bravais} lattice but that is beyond the purpose of this work.} in the limit $d \rightarrow \infty$~\cite{mu.we.18} with the usual rescaling of the hopping $t_{\text{c}}=t^{\ast}/(2\sqrt{d})$. Every momentum-dependent function then depends on the electronic crystal momemtum $\vec{k}$ only {\em via} $\epsilon = -2t_{\text{c}} \sum_{i=1}^{d} \cos(k_i a)$ and $\overline{\epsilon} = -2t_{\text{c}}\sum_{i=1}^{d} \sin(k_i a)$. Sums over the Brillouin zone are then performed using the joint density of states~\cite{tu.fr.05,ts.ok.08,ma.ga.22} $\rho(\epsilon,\overline{\epsilon}) = 1/(\pi t^{\ast 2}) \ \exp[-( \epsilon^{2} + \overline{\epsilon}^{2})/t^{\ast 2}]$.

Disorder is introduced via the site-dependent shifts of the onsite energies $V_{i}$ which are distributed uniformly according to the probability distribution function (PDF)
\begin{equation}\label{eq:disorder_PDF}
P(V)=\frac{1}{2W}\theta\left( W-|V| \right),
\end{equation}
$W$ being the disorder amplitude.

In this work we address disorder effects either by means of the CPA~\cite{do.te.22,ya.we.23u}, or making use of the SCB scheme~\cite{mu.ec.18}, which will be discussed in Sec.~\ref{sec:CPA_Floquet_appox}. In the regime in which CPA and SCB coincide there is no loss of generality in choosing a uniform PDF as in Eq.~\eqref{eq:disorder_PDF} since the SCB approximation does not depend on the impurities' distribution~\cite{ha.ja}.

Following Refs.~\cite{mu.we.15,mu.ts.17,ma.we.23} we couple each lattice site to an optical phonon branch of frequency $\omega_{\text{\tiny E}}$. The local e-ph interaction Hamiltonian then reads 
\begin{equation}\label{eq:e-ph_Ein_ham}
 \hat{H}_{\text{e-ph}} = g \sum_{i\sigma} \hat{n}^{f}_{i\sigma} \hat{x}_{i},
\end{equation}
where the phonon displacement operator $\hat{x}_{i}\equiv (\hat{b}^{\dagger}_{i} + \hat{b}_{i})/\sqrt{2}$ interacts with the electron density $\hat{n}^{f}_{i\sigma}$ through the e-ph coupling $g$. The operator $\hat{b}^{\dagger}_{i}$ ($\hat{b}_{i}$) creates (annihilates) an optical phonon with frequency $\omega_{\text{\tiny E}}$ at the lattice site $i$. The free phonon Hamiltonian consists of an Einstein phonon $\hat{H}_{\text{ph}} = \omega_{\text{\tiny E}}\sum_{i}\hat{n}^{b}_{i}$ with $\hat{n}^{b}_{i}=\hat{b}^{\dagger}_{i}\hat{b}_{i}$ the phonon density, coupled to an ohmic bath $\hat{H}_{\text{ph},\text{ohm}}$, the details of which will be given in Sec.~\ref{sec:e-ph_SE_impl}, see Eqs~\eqref{eq:ohmic_bath_GF} and~\eqref{eq:ohm_bath_spec}.
 
We set $\hbar = k_{\text{B}} = a = 1 = -Q$ so that $\Omega\equiv F$ and the current is then measured in units of the hopping $t^{\ast}$.
\section{Methods}\label{sec:method}

\subsection{Floquet electron Dyson equation}\label{sec:Dyson-eq}

Here we briefly introduce the non-equilibrium Floquet Green's function (GF) formalism~\cite{so.do.18,ma.ga.22,ga.ma.22,ma.we.23}: a more detailed derivation can be found in Ref.~\cite{ts.ok.08}.

We denote Floquet-represented matrices by either $X_{mn}$ or $\mat{X}$ (see, e.g. Refs.~\cite{so.do.18,ma.ga.22,ma.we.23}), while an underline refers to the Keldysh structure
\begin{equation}\label{eq:Keld-structure}
\kel{\mat{X}} \equiv 
\begin{pmatrix}
\mat{X}^{\text{R}} & \mat{X}^{\text{K}} \\
\mat{0}            & \mat{X}^{\text{A}} \\
\end{pmatrix}
\end{equation}
with $\mat{X}^{\text{R},\text{A},\text{K}}$ the {\em retarded}, {\em advanced} and {\em Keldysh} components obeying the usual relations $\mat{X}^{\text{A}}=(\mat{X}^{\text{R}})^{\dagger}$ and $\mat{X}^{\text{K}} \equiv \mat{X}^{>} + \mat{X}^{<}$, where $\mat{X}^{\lessgtr}$ are the \emph{lesser} and \emph{greater} components~\cite{schw.61,keld.65,ra.sm.86,ha.ja}.

Due to the presence of disorder the system is no longer translation invariant~\cite{ha.ja}, so that prior to averaging over the disorder configurations both the GF and self-energy (SE) depend on two {\em different} vectors of the Brillouin zone (BZ)~\cite{ab.dz.75}, namely $\kel{\vec{G}}(\omega,\vec{k},\vec{k}^{\prime})$ and $\kel{\vec{\Sigma}}_{\text{\tiny U}+\text{dis}}(\omega,\vec{k},\vec{k}^{\prime})$. In contrast to the standard notation used in the literature~\cite{ha.ja,ab.dz.75}, we do not introduce an additional symbol for {\em disorder-averaged} quantities. We simply denote such objects to be dependent only on one vector $\vec{k}$ of the BZ. The information contained in $\vec{k}$ is then {\em shifted} into the global variables~\cite{ts.ok.08} $\epsilon$, $\overline{\epsilon}$ which have been introduced in Sec.~\ref{sec:MO_HA}. As a result $\kel{\vec{G}}(\omega,\epsilon,\overline{\epsilon})$ and $\kel{\vec{\Sigma}}_{\text{\tiny U}+\text{dis}}(\omega,\epsilon,\overline{\epsilon})$ denote the disorder-averaged {\em interacting} GF and SE respectively.

It is worth mentioning that, in contrast to Refs.~\cite{ma.ga.22,ma.we.23}, in this work the electron SE $\kel{\vec{\Sigma}}_{\text{\tiny U}+\text{dis}}$ accounts for both electronic correlation and disorder-related effects. By setting $V_{i}=0$ in Eq.~(\ref{eq:Hubbard_ham}) one discards the disorder and recovers the setups described in~\cite{ma.ga.22,ma.we.23}.

For details about the two approximation hereby used to deal with disorder see Secs.~\ref{sec:CPA_Floquet_appox} and~\ref{sec:CPA_FDMFT_treatment}.

In presence of disorder, the Dyson equation for the electronic lattice GF of the system reads~\cite{so.do.18,ma.ga.22,ma.we.23}
\begin{equation}\label{eq:FullDysonEq}
\kel{\mat{G}}^{-1}(\omega,\epsilon,\overline{\epsilon}) = \kel{\mat{G}}^{-1}_{0}(\omega,\epsilon,\overline{\epsilon}) - \kel{\vec{\Sigma}}_{\text{\tiny U}+\text{dis}}(\omega,\epsilon,\overline{\epsilon}) - \kel{\mat{\Sigma}}_{\text{e-ph}}(\omega,\epsilon,\overline{\epsilon}),
\end{equation}
where in view of the Migdal approximation the self-energies $\kel{\vec{\Sigma}}_{\text{\tiny U}+\text{dis}}$ and $\kel{\mat{\Sigma}}_{\text{e-ph}}$ contribute separately.
In principle, both electron and e-ph self-energies depend on the electron crystal momentum via $\epsilon$, $\overline{\epsilon}$. However, due to the DMFT approximation we consider only local contributions, i.e. $\kel{\vec{\Sigma}}_{\text{\tiny U}+\text{dis}}(\omega,\epsilon,\overline{\epsilon}) \approx \kel{\vec{\Sigma}}_{\text{\tiny U}+\text{dis}}(\omega)$ and $\kel{\mat{\Sigma}}_{\text{e-ph}}(\omega,\epsilon,\overline{\epsilon}) \approx \kel{\mat{\Sigma}}_{\text{e-ph}}(\omega)$. The {\em noninteracting} GF $\kel{\mat{G}}_{0}$ in Eq.~\eqref{eq:FullDysonEq} describes the system when both electronic correlation and disorder are discarded ($U=0$ and $V_{i}=0$) and, as such, is always translation invariant.

The Floquet matrix elements of the electron GF of the noninteracting Hamiltonian~\eqref{eq:MicroHamiltonian} read
\begin{align}\label{eq:non-int_InvGF}
    \begin{split}
    [G_{0}^{-1}(\omega,\epsilon,\bar{\epsilon})]^{\text{R}}_{mn} & = \left[ \omega_n-\varepsilon_c - \Sigma^{\text{R}}_{\text{bath}}(\omega_n) \right]\delta_{mn} - \varepsilon_{mn}(\epsilon,\overline{\epsilon}), \\
    [G_{0}^{-1}(\omega,\epsilon,\bar{\epsilon})]^{\text{K}}_{mn} & = - \delta_{mn} \Sigma^{\text{K}}_{\text{bath}}(\omega_{n})
    \end{split}
\end{align}
with the shorthand notation $\omega_{n}\equiv \omega+nF$ and $\varepsilon_{mn}(\epsilon,\overline{\epsilon})$ being the {\em Floquet dispersion relation}~\cite{ts.ok.08}
\begin{equation}\label{eq:Floquet_disp}
\varepsilon_{mn}(\epsilon,\overline{\epsilon}) = \frac{1}{2} \left[ \left( \epsilon + \ii \overline{\epsilon} \right)\delta_{m-n,1} +  \left( \epsilon - \ii \overline{\epsilon} \right)\delta_{m-n,-1} \right].
\end{equation}
Following Refs.~\cite{aron.12,ne.ar.15,ma.ga.22,ma.we.23}, the fermionic heat bath $\hat{H}_{\text{bath}}$ contributes the SE
\begin{align}\label{eq:WBL_bathGF}
\begin{split}
\Sigma^{\text{R}}_{\text{bath}}(\omega) & = - \ii \Gamma_{\text{e}}/2 \\
\Sigma^{\text{K}}_{\text{bath}}(\omega) & = -\ii \Gamma_{\text{e}} \tanh \left[ \beta\left(\omega-\mu\right)/2\right]
\end{split}
\end{align}
to the noninteracting GF~\eqref{eq:non-int_InvGF}, where $\Gamma_{\text{e}}$ defines the electronic dephasing rate~\footnote{We recall that the {\em Keldysh} component in Eq.~(\ref{eq:WBL_bathGF}) is obtained from the {\em fluctuation-dissipation theorem}, i.e. $\Sigma^{\text{K}}_{\text{bath}}(\omega) = 2\ii \text{Im}[\Sigma^{\text{R}}_{\text{bath}}(\omega)] \tanh \left[ \beta\left(\omega-\mu\right)/2\right]$, since the heat bath is at equilibrium.}. The inverse temperature and chemical potential of the bath are denoted by $\beta$ and $\mu$ respectively.

\subsection{SCB scheme}\label{sec:CPA_Floquet_appox}

In this section we introduce the SCB approximation~\cite{ha.ja} to the disorder, the SE of which is given by
\begin{equation}\label{eq:SC_Born_SE}
 \kel{\mat{\Sigma}}_{\text{\tiny SCB}}(\omega,\vec{k}) =  \sum_{\vec{k}^{\prime}} \overline{|V(\vec{k}-\vec{k}^{\prime})|^{2}} \kel{\mat{G}}(\omega,\vec{k}^{\prime}),
\end{equation}
where $\overline{\cdots}$ denotes disorder averaging. 

We can then disentagle the disorder effects from electronic correlation, i.e.
\begin{equation}\label{eq:towards_DIA}
 \kel{\mat{\Sigma}}_{\text{\tiny U}+\text{dis}}(\omega,\vec{k}) = \kel{\vec{\Sigma}}_{\text{\tiny U}}(\omega) + \kel{\mat{\Sigma}}_{\text{\tiny SCB}}(\omega,\vec{k}),
\end{equation}
where we made use of the DMFT approximation to take a local electron SE $\kel{\vec{\Sigma}}_{\text{\tiny U}}(\omega)$.

\begin{figure*}[t]
  \includegraphics[width=0.8\linewidth,keepaspectratio]{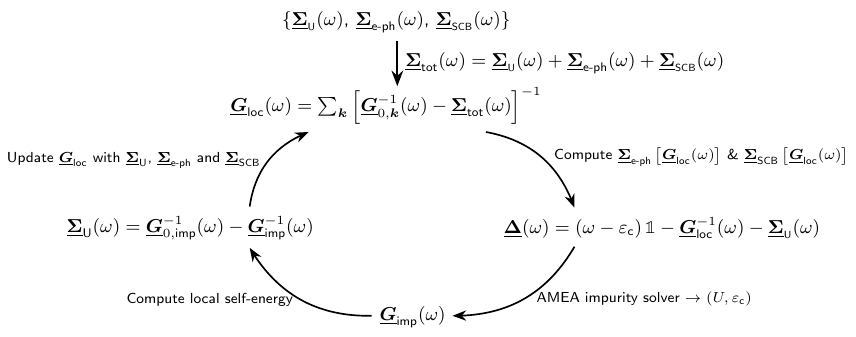}
  \caption{Flowchart of the DMFT loop within the $\text{SCB}$ scheme. Notice that the electron bath SE $\kel{\mat{\Sigma}}_{\text{bath}}$, see Eq.~\eqref{eq:non-int_InvGF}, is contained in the noninteracting {\em disorder-averaged} lattice GF $\kel{\mat{G}}_{0,\vec{k}}(\omega)$: for the sake of simplicity we restored the notation $\vec{k}$ to denote the electron crystal momentum, see also Sec.~\ref{sec:Dyson-eq}. The e-ph SE $\vec{\kel{\Sigma}}_{\text{e-ph}}(\omega)$ is computed by Fourier-transforming Eq.~\eqref{eq:backbone_e-ph_SE} and the SCB SE $\kel{\mat{\Sigma}}_{\text{\tiny SCB}}(\omega)$ is obtained as in Eq.~\eqref{eq:DIA_SE}.}
\label{fig:FDMFT_DIA_scheme}
\end{figure*}

In our case disorder consists of impurities with local potentials, so that the Fourier transform of $V$ is $k$-independent, thus leading to
\begin{equation}\label{eq:DIA_SE}
\kel{\mat{\Sigma}}_{\text{\tiny SCB}}(\omega) = \overline{V^{2}} \kel{\mat{G}}_{\text{loc}}(\omega),
\end{equation}
In Eq.~\eqref{eq:DIA_SE}, $\kel{\mat{G}}_{\text{loc}}$ is the {\em disorder-averaged} local electron GF
 \begin{equation}\label{eq:Lat_LocGF}
 \kel{\mat G}_{\text{loc}}(\omega) = \int \dd\epsilon \int \dd\overline{\epsilon} \ \rho(\epsilon,\overline{\epsilon}) \left[ \kel{\mat G}^{-1}_{0}(\omega,\epsilon,\overline{\epsilon}) -\kel{\mat\Sigma}_{\text{tot}}(\omega) \right]^{-1}, 
 \end{equation}
where
\begin{equation}\label{eq:DIA_tot_SE}
\kel{\mat\Sigma}_{\text{tot}}(\omega) = \kel{\mat\Sigma}_{\text{\tiny U}}(\omega) + \kel{\mat \Sigma}_{\text{e-ph}}(\omega) + \kel{\mat{\Sigma}}_{\text{\tiny SCB}}(\omega)
\end{equation}
is the total SE, see also the flowchart in Fig.~\ref{fig:FDMFT_DIA_scheme}.

Notice that the combination in square brackets in Eq.~\eqref{eq:Lat_LocGF} corresponds to the full Dyson equation~\eqref{eq:FullDysonEq} with the addition of $\kel{\mat{\Sigma}}_{\text{\tiny SCB}}(\omega)$: by virtue of the DMFT and SCB approximations all the self-energies are now local. 

To conclude this section we briefly recall the basic concepts about the $\text{DMFT}$~\footnote{For details about the DMFT loop we point at Refs.~\cite{so.do.18,ma.ga.22,ga.ma.22,ma.we.23}.}, which, even though not specific to the SCB approximation will show how the DMFT loop is modified in presence of disorder within the SCB approach and will serve as reference for the CPA scheme.

The basic idea is to map the lattice problem onto a single-site {\em impurity model} accounting for the remaining sites through the {\em hybridization} function $\kel{\vec{\Delta}}(\omega)$, which has to be self-consistently determined from
\begin{equation}\label{eq:imp_DFMT_equivalence}
 \kel{\vec{G}}^{-1}_{\text{imp}}(\omega) = \kel{\vec{g}}^{-1}_{0,\text{site}}(\omega) - \kel{\vec{\Delta}}(\omega) - \kel{\vec{\Sigma}}_{\text{\tiny U}}(\omega).
\end{equation}
by requiring $\kel{\mat{G}}_{\text{imp}}(\omega) \overset{!}{=} \kel{\mat{G}}_{\text{loc}}(\omega)$, with $\kel{\mat{G}}_{\text{loc}}(\omega)$ as in Eq.~\eqref{eq:Lat_LocGF}. In Eq.~\eqref{eq:imp_DFMT_equivalence} $\kel{\vec{g}}_{0,\text{site}}$ denotes the noninteracting impurity GF, the retarded component of which reads $\kel{\vec{g}}^{-1,\text{R}}_{0,\text{site}} = \omega - \varepsilon_{\text{c}}$. In practice, after starting with a suitable guess for the self-energies $\kel{\vec{\Sigma}}_{\text{\tiny U}}(\omega)$, $\kel{\mat{\Sigma}}_{\text{e-ph}}(\omega)$ and $\kel{\mat{\Sigma}}_{\text{\tiny SCB}}(\omega)$, we compute $\kel{\vec{\Delta}}(\omega)$ according to Eq.~\eqref{eq:imp_DFMT_equivalence} and use it to generate~\footnote{For further details about the $\text{DMFT}$ loop we refer to~\cite{ma.ga.22} (in particular see Sec. III C therein) and to the flowchart in Fig.~\ref{fig:FDMFT_DIA_scheme} in this Manuscript, while the latest developments concerning the AMEA impurity solver have been discussed in Refs~\cite{we.lo.23,ma.we.23}.} the new electron SE $\kel{\vec{\Sigma}}_{\text{\tiny U}}(\omega)$. We then iterate this procedure until convergence is reached. We stress that due to the time-translation invariance of the problem at hand, a local equation of the form~\eqref{eq:imp_DFMT_equivalence} can be restricted to the diagonal matrix elements only. Due to the property of the Floquet matrices~\cite{ts.ok.08}, the emergent NESS is then characterized by the $(0,0)$-component alone, see Appendix~\ref{sec:GFs_Dyson_Floquet} for further details~\footnote{The size of the matrices in the Floquet sector is $N_{\text{F}}=21$ and it has been chosen in such a way that the electron features are converged with respect to it.}.

\subsection{CPA scheme}\label{sec:CPA_FDMFT_treatment}

The CPA treatment~\cite{jani.89,ja.vo.92,do.te.22,ya.we.23u} consists in taking the SE obtained from the {\em disorder-averaged} local GF as the SE of the disorder-averaged GF of the system.

As we mentioned in Sec.~\ref{sec:intro}, the $\text{CPA}$ scheme~\cite{do.te.22,ya.we.23u} requires the solution of several impurity problems corresponding to the {\em shifted} onsite energies
\begin{equation}\label{eq:shifted_onsite_ngr}
 \varepsilon(V)\equiv\varepsilon_{\text{c}}+V,
\end{equation}
see also the flowchart in Fig.~\ref{fig:FDMFT_CPA_scheme}. Solving the impurity problem for all the inequivalent configurations then yields a set of $V$-dependent self-energies $\lbrace \kel{\mat{\Sigma}}_{\text{\tiny U+dis}}(\omega,V)\rbrace$, or equivalently impurity Greeen's functions $\lbrace \kel{\mat{G}}_{\text{imp}}(\omega,V)\rbrace$, solutions to the impurity problems {\em uniquely} identified by $V$ and $U$~\footnote{We recall that in the particle-hole symmetric case considered in this work $\varepsilon_{\text{c}}=-U/2$.}. One then defines the {\em disorder-averaged} impurity GF as
\begin{equation}\label{eq:exact_FDMFT_CPA}
   \langle \kel{\mat{G}}_{\text{imp}}(\omega) \rangle_{\left\{ V \right\}} \equiv \int \dd V P(V) \kel{\mat{G}}_{\text{imp}}(\omega,V),
\end{equation}
where the disorder PDF $P(V)$ is given in Eq.~\eqref{eq:disorder_PDF}. For a particle-hole symmetric disorder distribution $P(V)$ the disorder-averaged impurity GF $\langle \kel{\mat{G}}_{\text{imp}}(\omega) \rangle_{\left\{ V \right\}}$ obviously preserves particle-hole symmetry and the same holds for all the other quantities such as $\kel{\vec{\Delta}}(\omega)$ and $\kel{\vec{\Sigma}}_{\text{\tiny U+dis}}(\omega)$.

The workflow then goes as follows: once $\langle \kel{\mat{G}}_{\text{imp}}(\omega) \rangle_{\left\{ V \right\}}$ in Eq.~\eqref{eq:exact_FDMFT_CPA} is computed we use it to extract the new SE $\kel{\vec{\Sigma}}_{\text{\tiny U+dis}}$, see the flowchart in Fig.~\ref{fig:FDMFT_CPA_scheme}. We then add to it the e-ph SE $\kel{\vec{\Sigma}}_{\text{e-ph}}$ and insert them into Eq.~\eqref{eq:Lat_LocGF}. By requiring $\langle \kel{\mat{G}}_{\text{imp}}(\omega) \rangle_{\left\{ V \right\}} \overset{!}{=} \kel{\mat{G}}_{\text{loc}}(\omega)$ the hybridization $\kel{\vec{\Delta}}$ can be extracted from
\begin{equation}\label{eq:extract_Delta}
    \langle \kel{\mat{G}}_{\text{imp}}(\omega) \rangle^{-1}_{\left\{ V \right\}} = \kel{\vec{g}}^{-1}_{0,\text{site}}(\omega) - \kel{\vec{\Delta}}(\omega) - \kel{\vec{\Sigma}}_{\text{\tiny U+dis}}(\omega),
\end{equation}
and the DMFT loop proceeds as described in Sec.~\ref{sec:CPA_Floquet_appox}.

\subsection{Electron-phonon interaction}\label{sec:e-ph_SE_impl}

Within DMFT the e-ph SE is taken to be local, namely $\kel{\mat{\Sigma}}_{\text{e-ph}}(\omega,\epsilon,\overline{\epsilon}) \approx \kel{\mat{\Sigma}}_{\text{e-ph}}(\omega)$. In terms of the {\em contour-times} $z,z^\prime$, and in the Migdal approximation~\footnote{Notice that in the Migdal approximation the {\em Hartree term} amounts to a constant energy shift that can be reabsorbed in the interacting e-ph Hamiltonian $\hat{H}_{\text{e-ph}}$ at half-filling.}, $\kel{\mat{\Sigma}}_{\text{e-ph}}$ reads~\cite{mu.we.15,mu.ts.17,ma.ga.22,ma.we.23}
\begin{equation}\label{eq:backbone_e-ph_SE}
\Sigma_{\text{e-ph}}(z,z^{\prime}) = \ii g^{2} G_{\text{loc}}(z,z^{\prime}) D_{\text{ph}}(z,z^{\prime}),
\end{equation}
where $G_{\text{loc}}(z,z^{\prime})$ is the {\em contour-times} local electron GF allowing the representation in Eq.~\eqref{eq:Lat_LocGF} in frequency domain. Optical phonons are quickly discussed below.

\subsubsection{Phonon Dyson equation}\label{sec:Ph_Dyson}

Following the derivation in Refs~\cite{mu.ts.17,ma.we.23} we model the optical phonon branch by Einstein phonons coupled to an ohmic bath, the Dyson equation of which reads
\begin{equation}\label{eq:local_Dyson_Ein_ph}
\kel{D}_{\text{ph}}(\omega) = [\kel{D}^{-1}_{\text{ph},\text{\tiny E}}(\omega) - \kel{\Pi}_{\text{ohm}}(\omega) - \kel{\Pi}_{\text{e-ph}}(\omega)]^{-1}
\end{equation}
with the {\em retarded} non-interacting Einstein phonon propagator~\footnote{As usual, the Keldysh component $D^{\text{K}}_{\text{ph}}(\omega)$ can be neglected due to the presence of the ohmic bath $\kel{\Pi}_{\text{ohm}}$ in Eq.~(\ref{eq:local_Dyson_Ein_ph}).} given by
\begin{align}\label{eq:non-int_einstein_ph}
D^{\text{R}}_{\text{ph},\text{\tiny E}}(\omega) = \frac{2\omega_{\text{\tiny E}}}{\omega^{2} - \omega_{\text{\tiny E}}^{2}}.
\end{align}

The Einstein phonon is coupled to an ohmic bath $\hat{H}_{\text{ph},\text{ohm}}$, the real {\em retarded} GF of which is obtained from the Kramers-Kr\"onig relations (see e.g. Ref.~\cite{mu.ts.17}), having the following {\em Keldysh} component
\begin{equation}\label{eq:ohmic_bath_GF}
\Pi^{\text{K}}_{\text{ohm}}(\omega) = -2\pi\ii A_{\text{ohm}}(\omega) \coth(\beta\omega/2).
\end{equation}
The ohmic bath DOS in~\eqref{eq:ohmic_bath_GF} is taken as
\begin{equation}\label{eq:ohm_bath_spec}
A_{\text{ohm}}(\omega) = \frac{v^{2}_{\text{o}}}{\omega_{\text{o}}} \left[ \frac{1}{1+\left( \frac{\omega-\omega_{\text{o}}}{\omega_{\text{o}}}\right)^{2}}  - \frac{1}{1+\left( \frac{\omega+\omega_{\text{o}}}{\omega_{\text{o}}}\right)^{2}} \right]
\end{equation}
with the usual definition $-\pi A_{\text{ohm}}(\omega) \equiv \text{Im}[\Pi^{\text{R}}_{\text{ohm}}(\omega)]$. In Eq.~\eqref{eq:ohm_bath_spec} $\omega_{\text{o}}$ denotes the ohmic bath cutoff frequency and $v_{\text{o}}$ the hybridization strength to the ohmic bath~\footnote{Notice that Eq.~(\ref{eq:ohm_bath_spec}) ensures a linear dependence within almost the entire interval $\omega \in [-\omega_{\text{o}},\omega_{\text{o}}]$.}.

\subsubsection{Phonon self-energy}\label{sec:self-cons_phonons}

Within the Migdal approximation, the contour times phonon self-energy~\cite{mu.we.15,mu.ts.17} in Eq.~\eqref{eq:local_Dyson_Ein_ph} reads
\begin{align}\label{eq:bubble_GG}
\Pi_{\text{e-ph}}(z,z^{\prime})=-2\ii g^{2} G_{\text{loc}}(z,z^{\prime})G_{\text{loc}}(z^{\prime},z)
\end{align}
with $G_{\text{loc}}(z,z^{\prime})$ being the local electron GFs on the Keldysh contour allowing the representation~\eqref{eq:Lat_LocGF} and the factor $2$ accounts for spin degeneracy. The real time components of Eq.~\eqref{eq:bubble_GG} have been derived in Ref.~\cite{ma.we.23} (see Appendix C therein).

\subsection{Observables}\label{sec:observables}

\begin{figure*}[t]
\includegraphics[width=0.8\linewidth,keepaspectratio]{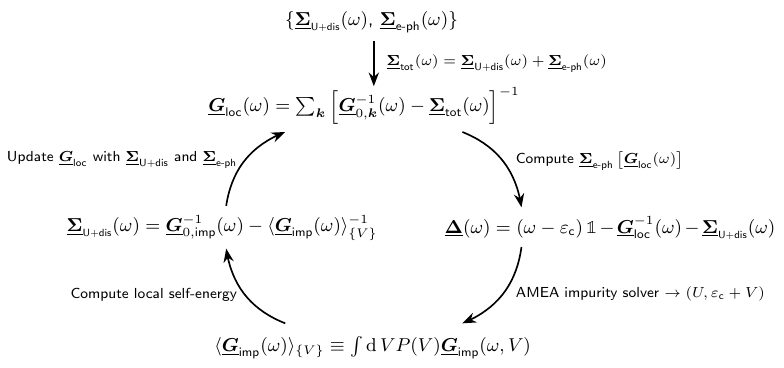}
\caption{Flowchart of the DMFT loop in the $\text{DMFT}+\text{CPA}$ scheme. The same considerations as in Fig.~\ref{fig:FDMFT_DIA_scheme} hold in this case too. Notice that $\kel{\mat{G}}_{0,\text{imp}}$ is the solution to the noninteracting impurity problem with {\em bare} onsite energy $\varepsilon_{\text{c}}$, see also Sec.~\ref{sec:CPA_FDMFT_treatment}.}
\label{fig:FDMFT_CPA_scheme}
\end{figure*}

Here we summarize the observables of interest in this Manuscript. We start from the {\em local} electron spectral function (SF)
\begin{align}\label{eq:local_spec_func}
 A(\omega) =-\text{Im}[G^{\text{R}}_{\text{loc}}(\omega)]/\pi,
\end{align}
which can be used to define the {\em local} electron spectral occupation function as
 \begin{equation}\label{eq:Filling_func}
 N_{\text{e}}(\omega) \equiv A(\omega)\left\{ \frac{1}{2} - \frac{1}{4} \frac{\text{Im}[G^{\text{K}}_{\text{loc}}(\omega)]}{\text{Im}[G^{\text{R}}_{\text{loc}}(\omega)]} \right\},
 \end{equation}
 where the expression in curly brackets is the {\em local} electron non-equilibrium distribution function
 \begin{equation}\label{eq:NEFD-dist}
 F_{\text{el}}(\omega) \equiv \frac{1}{2} \left\{1 - \frac{1}{2}\frac{\text{Im}[G^{\text{K}}_{\text{loc}}(\omega)]}{\text{Im}[G^{\text{R}}_{\text{loc}}(\omega)]} \right\}.
 \end{equation}

In our units, the steady-state current~\footnote{As pointed out in Refs~\cite{ts.ok.08,ma.ga.22,ma.we.23}, any quantity with a single index refers to the Wigner representation, which obeys the relation $\kel{X}_{mn}(\omega)=\kel{X}_{m-n}(\omega+(m+n)\Omega/2)$, where $X_{mn}(\omega)$ is any Floquet-represented matrix.} flowing along the direction of the applied field~\cite{ma.ga.22,ma.we.23} is given by
\begin{equation}\label{eq:general_Wig_current}
  J = \int \frac{\dd\omega}{2\pi} \int \dd\epsilon \int \dd\overline{\epsilon} \ \rho(\epsilon,\overline{\epsilon}) \left[ \left( \epsilon - \ii \overline{\epsilon} \right) G^{<}_{1}(\omega,\epsilon,\overline{\epsilon}) + \text{H.c.} \right].
\end{equation}

\section{Results}\label{sec:General_Results}

\begin{table}[t]
  \begin{center}
\begin{tabular}{ cccccccccc }
      \hline
      \hline
        $U$ \ & $\varepsilon_{\text{c}}$ \ & $\mu$ \ & $1/\beta$ \ & $\overline{V^{2}}$ \ & $g$ \ & $\omega_{\text{\tiny E}}$ \ & $v_{\text{o}}$ \ & $\omega_{\text{o}}$ \\
      \hline
        $8$ \ & $-4$ \ & $0$ \ & $0.05$ \ & $0.083$ \ & $0.4$ \ & $0.6$ \ & $0.055$ \ & $0.6$ \\
      \hline
      \hline
    \end{tabular}
    \caption{Default values of the main parameters used in this Manuscript. All parameters are given in units of $t^{\ast}$.}
    \label{tab:default_pars}
  \end{center}
\end{table}

\subsection{Benchmarking $\text{SCB}$ against $\text{CPA}$}\label{sec:CPA_vs_DIA_benchmark}

\begin{figure}[b]
 \includegraphics[width=0.9\linewidth]{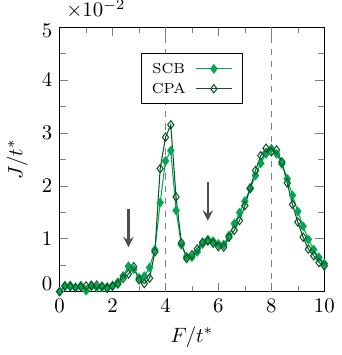}
 \caption{Current $J$ as function of the applied field $F$ obtained within the SCB and CPA schemes. Black vertical arrows point at the resonances $F \approx U/3$ and $F \approx 2U/3$, while vertical dashed lines mark the position of the two main resonances $F=U/2$, $U$, the spectral functions of which are shown in Fig.~\ref{fig:DIA_CPA_spectra}. Default parameters can be found in Table~\ref{tab:default_pars}, with $W$ as in Eq.~\eqref{eq:DIA_CPA_dis_amplitudes}. (Here $U=8t^{\ast}$ and $\Gamma_{\text{e}}=6\times 10^{-2}t^{\ast}$.)}
 \label{fig:currents_DIA_CPA}
\end{figure}

As discussed in Sec.~\ref{sec:CPA_FDMFT_treatment} (see also the flowchart in Fig.~\ref{fig:FDMFT_CPA_scheme}) the CPA approximation is computationally costly due to the fact that several impurity problems have to be solved for each DMFT loop. For this reason we are interested in understanding in which circumstances the $\text{CPA}$ and $\text{SCB}$ schemes yield comparable results. In this section we then benchmark the two approximations against one another.

To do so we notice that by averaging the square of the perturbations $V_{i}$'s according to the PDF in Eq.~\eqref{eq:disorder_PDF} within the CPA scheme we get
\begin{equation}
\langle V^{2} \rangle \equiv \int \dd V P(V) V^{2} = \frac{W^{2}}{3},
\end{equation}
which can be interpreted as the square of the {\em effective} disorder amplitude in the SCB scheme (see Eq.~\eqref{eq:DIA_SE}), i.e. in the regime in which CPA can be replaced by SCB we can identify
\begin{equation}\label{eq:DIA_CPA_dis_amplitudes}
\overline{V^{2}}=W^{2}/3.
\end{equation}

\subsubsection{Current characteristics}\label{sec:dia_cpa_curr_bench}

We start by comparing the current-field characteristics.
By matching the parameters $W$ and $\overline{V^{2}}$ according to Eq.~\eqref{eq:DIA_CPA_dis_amplitudes} the $J$-$F$ curves in the CPA and SCB schemes lie almost on top of each other, see Fig.~\ref{fig:currents_DIA_CPA}. In particular one can see that the current $J$ takes on comparable values in the two approaches for almost all the values of the applied field. 
In particular the two currents coincide at the two resonances $F\approx U/3$, $2U/3$, highlighted by black arrows in Fig.~\ref{fig:currents_DIA_CPA}, and at $F=U$. The most pronounced difference occurs for field strengths roughly equal to half of the band gap, i.e. $F\approx U/2$: at around this resonance we see that the current $J$ takes on larger values in the CPA scheme as opposed to the SCB approach, see again Fig.~\ref{fig:currents_DIA_CPA}.
An interpretation of these differences in terms of the spectral function is presented below~\footnote{To conclude we just want to mention that the region $F/t^{\ast} \in [0,1]$ required a higher resolution in the AMEA impurity solver as the {\em hybridization} functions corresponding to those field strengths contain very fine features which, if not correctly resolved, could lead to artefacts in the observables. For a detailed discussion about the impurity solver hereby employed we refer to our recent work~\cite{we.lo.23,ma.we.23}.}.

\begin{figure}[b]
 \includegraphics[width=\linewidth]{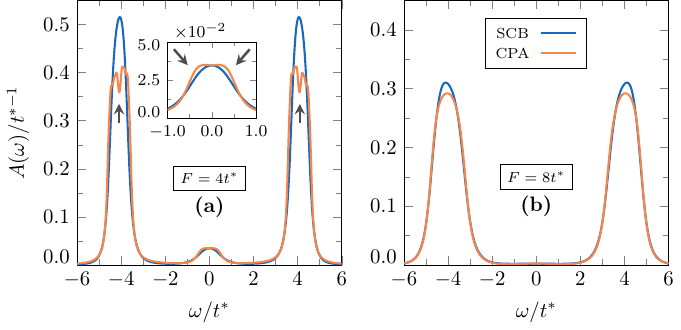}
 \caption{(a) Electron SF $A(\omega)$ at $F=4t^{\ast}$ within the SCB and CPA schemes. Black arrows highlight the {\em kinks} occurring in the spectral features in the CPA approach. The inset magnifies the region $\omega\approx 0$, with black arrows pointing at the broadening of the spectrum in the CPA scheme. (b) Same quantity for $F=8t^{\ast}$. Default parameters can be found in Table~\ref{tab:default_pars}, with $W$ as in Eq.~\eqref{eq:DIA_CPA_dis_amplitudes}. (Here $U=8t^{\ast}$ and $\Gamma_{\text{e}}=6\times 10^{-2}t^{\ast}$.)}
 \label{fig:DIA_CPA_spectra}
\end{figure}
\subsubsection{Spectral features}
After having shown that the $J$-$F$ curves coincide over almost the whole range of field strengths of interest, we now compare the electronic SF around the resonances $F=U/2$ and $F=U$ to characterize the most important current-carrying regimes. In particular, at $F=U/2$ the CPA $A(\omega)$ exhibits a broader {\em quasi-particle peak} (QPP) at $\omega\approx 0$ as opposed to the SCB case, see Fig.~\ref{fig:DIA_CPA_spectra}(a) and corresponding inset. In addition, the CPA spectral weight at $\omega \approx \pm U/2$ is reduced, with some {\em kinks} occurring in the main Hubbard bands, highlighted by the black vertical arrows in Fig.~\ref{fig:DIA_CPA_spectra}(a). 

Most likely the differences in the current characteristics shown in Fig.~\ref{fig:currents_DIA_CPA} at $F\approx U/2$ can be attributed to the {\em qualitative} change in $A(\omega)$ between the two approaches, see again Fig.~\ref{fig:DIA_CPA_spectra}(a). Such differences become only {\em quantitative} for fully resonant fields $F=U$, while the overall structure of $A(\omega)$ is preserved, see Fig.~\ref{fig:DIA_CPA_spectra}(b). This is due to the fact that at $F=U$ the occupations $N_{\text{e}}(\omega)$ of the upper Hubbard band (UHB) in the two approaches (not shown) are essentially the same, so that the difference in the SF becomes less relevant for the $J$-$F$ characteristics.

In conclusion, having shown that the CPA and SCB schemes yield comparable results as far as the conducting properties are concerned~\footnote{Of course, this is valid only for corresponding values of the parameters as identified by Eq.~(\ref{eq:DIA_CPA_dis_amplitudes}).}, we can then employ the latter to investigate the role played by disorder in Mott insulating systems, as it is computationally cheaper.

\subsection{Interacting disordered system without phonons}

In this section we study the effects of disorder on the conducting properties of the system at hand by comparing the setups $\overline{V^{2}}=0$ and $\overline{V^{2}}=0.083t^{\ast 2} = W^{2}/3$, the latter being the value corresponding to a disorder distribution given by Eq.~\eqref{eq:disorder_PDF}, cf. the discussion in Sec.~\ref{sec:CPA_vs_DIA_benchmark}.

\subsubsection{Current characteristics and charge excitation}\label{sec:curr_charge_no_ph}

\begin{figure}[b]
 \includegraphics[width=\linewidth]{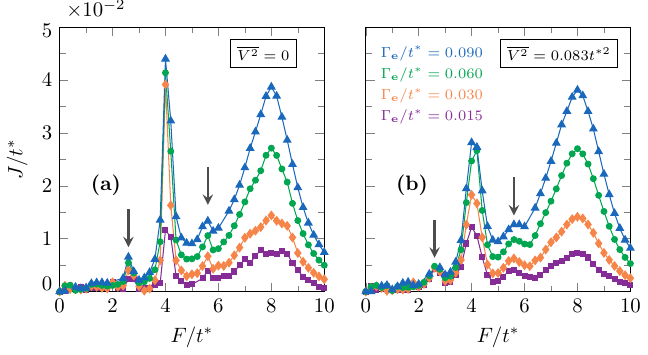}
 \caption{Current $J$ as function of the applied field $F$ at selected values of the dephasing rate $\Gamma_{\text{e}}$  (a) without ($\overline{V^{2}}=0$) and (b) with disorder ($\overline{V^{2}}=0.083t^{\ast 2}$). Black vertical arrows point at the resonances $F \approx U/3$ and $F \approx 2U/3$. The values of the dephasing rate $\Gamma_{\text{e}}$ are specified in the plot, while default parameters can be found in Table~\ref{tab:default_pars}. When present, disorder is treated in the SCB approximation. (Here $U=8t^{\ast}$.)}
 \label{fig:currents_DIA_Fs}
\end{figure}

Figure~\ref{fig:currents_DIA_Fs}(a) shows the current $J$ as function of the applied field $F$ for selected values of the electronic dephasing rate $\Gamma_{\text{e}}$ at $\overline{V^{2}}=0$. Here we observe a suppression of the current as $\Gamma_{\text{e}}$ is reduced for all the values of the electric field $F$, especially at the main resonances $F=U/3$, $U/2$ and $U$ and also at the one $F\approx 2U/3$~\cite{mu.we.18,ma.ga.22,ma.we.23}.
\begin{figure}[b]
 \includegraphics[width=\linewidth]{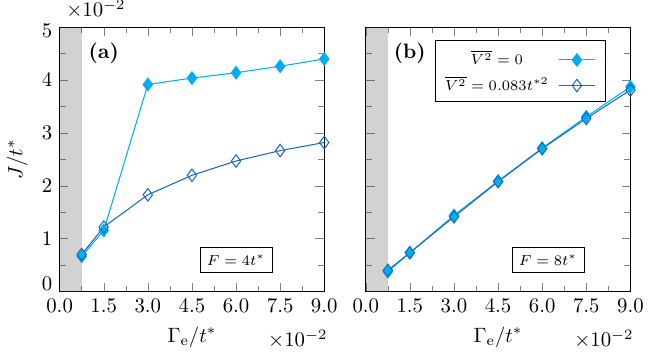}
 \caption{Current $J$ as function of the dephasing rate $\Gamma_{\text{e}}$ at selected values of $\overline{V^{2}}$ for applied fields (a) $F=4t^{\ast}$ and (b) $F=8t^{\ast}$. Gray shaded rectangles highlight the regions where the DMFT loop is unstable due to the very small dephasing rate $\Gamma_{\text{e}}$. Disorder amplitudes are specified in the plot, while default parameters are specified in Table~\ref{tab:default_pars}. When present, disorder is treated in the SCB approximation. (Here $U=8t^{\ast}$.)}
 \label{fig:currents_DIA_gammaes}
\end{figure}
The $J$-$F$ curve in presence of disorder is shown in Fig.~\ref{fig:currents_DIA_Fs}(b): we observe that all the main resonances are smeared out as opposed to the setup $\overline{V^{2}}=0$ shown in Fig.~\ref{fig:currents_DIA_Fs}(a). Also, in contrast to this case, the $J$-$F$ curve in presence of disorder looks {\em smoother} especially at $F=U/2$ and for small values of the dephasing rate $\Gamma_{\text{e}}$.

From Fig.~\ref{fig:currents_DIA_Fs}(b) we see that the reduction of the current with a smaller dephasing rate is a robust feature, occurring also in presence of disorder~\footnote{This result is valid for large fields where dissipation is essential in order to sustain a steady state current. Below a certain threshold field (not shown), we expect this behavior to be reversed.}. However, disorder alone is not expected to sustain a steady-state current as it provides only an elastic scattering mechanism~\cite{ha.ja}. In terms of the conducting properties this is evidenced by the dependence of $J$ on the dephasing rate $\Gamma_{\text{e}}$ at the two main resonances $F=U/2$ and $F=U$ with and without disorder. 

In Fig.~\ref{fig:currents_DIA_gammaes}(a) we plot the current $J$ as function of $\Gamma_{\text{e}}$ at $F=U/2$. We observe that in absence of disorder $J$ decreases slowly with decreasing $\Gamma_{\text{e}}$, until it abruptly drops when the dephasing rate is reduced below a threshold value $\Gamma^{\text{th}}_{\text{e}}=3\times 10^{-2}t^{\ast}$.~\footnote{We cannot reach down to $\Gamma_{\text{e}}=0$, as it is hard to get to a converged solution of the DMFT loop for too small values of $\Gamma_{\text{e}}$.}
On the other hand, in presence of disorder the current $J$ scales {\em sub-linearly} and it goes to zero smoothly as function of a decreasing $\Gamma_{\text{e}}$. However, the current $J$ takes on comparable values with and without disorder when the dephasing rate gets smaller than the threshold $\Gamma^{\text{th}}_{\text{e}}$ while it is suppressed for $\overline{V^{2}}=0.083t^{\ast 2}$ above $\Gamma^{\text{th}}_{\text{e}}$, see again Fig.~\ref{fig:currents_DIA_gammaes}(a). An understanding of these results in terms of the electronic spectral features will be given in Sec.~\ref{sec:results_dia_no_ph}.

\begin{figure}[t]
 \includegraphics[width=\linewidth]{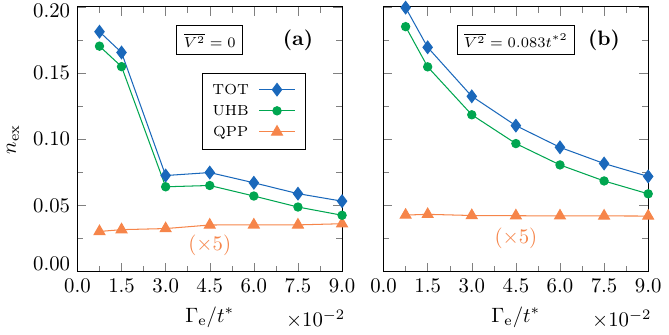}
 \caption{(a) Fraction of excited electrons $n_{\text{ex}}$ as function of the dephasing rate $\Gamma_{\text{e}}$ in absence of disorder at $F=4t^{\ast}$. (b) Same quantity in presence of disorder. $n_{\text{ex}}$ is obtained by integrating $N_{\text{e}}(\omega)$ in Eq.~\eqref{eq:Filling_func} over the whole positive energy spectrum (diamonds), around the UHB (dots) and the positive energy region of the QPP (triangles). Default parameters can be found in Table~\ref{tab:default_pars}. When present, disorder is treated in the SCB approximation. (Here $U=8t^{\ast}$.)}
 \label{fig:DIA_n_ex}
\end{figure}

In Fig.~\ref{fig:currents_DIA_gammaes}(b) the current $J$ is shown as function of the dephasing rate $\Gamma_{\text{e}}$ at $F=U$. For these values of $F$ compensating the band gap, disorder essentially has no effects on the current and its scaling behavior. Again, a more detailed discussion of the electron SF can be found in Sec.~\ref{sec:results_dia_no_ph}.

It is worth investigating the dependence of the fraction of excited electrons $n_{\text{ex}}$ around the positive region of the QPP ($\omega \geq 0$) and the UHB centered about $\omega \approx U/2$ in the metallic phase~\footnote{At $F=U/2$ we can indeed talk about metallic phase due to the occurrence of the maximum of in-gap states located at $\omega=0$ which provide the necessary spectral weight to sustain electron tunneling across the band gap.}, namely $F=U/2$. To obtain $n_{\text{ex}}$ we integrate the non-equilibrium occupation function $N_{\text{e}}(\omega)$ in Eq.~\eqref{eq:Filling_func} in the regions of the spectrum corresponding to the QPP ($\omega \geq 0$ only) and the UHB: the results with and without disorder effects are shown in Fig.~\ref{fig:DIA_n_ex}. As expected, in both cases the most important contribution to the fraction of excited electrons $n_{\text{ex}}$ comes from the occupation of the UHB since the in-gap states are only {\em bridging} the main bands. Accordingly, the occupation of the positive-energy states around the QPP barely depends on the dephasing rate $\Gamma_{\text{e}}$ with and without disorder, see Figs.~\ref{fig:DIA_n_ex}(a) and~\ref{fig:DIA_n_ex}(b). 

As pointed out in previous work, the dephasing rate $\Gamma_{\text{e}}$ is responsible for relaxation of high-energy electrons to the lower Hubbard band. In this context it is then clear that the UHB gets emptied the more by a larger dephasing rate, with and without disorder, see again Figs.~\ref{fig:DIA_n_ex}(a) and~\ref{fig:DIA_n_ex}(b). The fraction of electrons $n_{\text{ex}}$ in the UHB in absence of disorder shows the same {\em threshold-like} behavior as the current, see Figs.~\ref{fig:currents_DIA_gammaes}(a) and~\ref{fig:DIA_n_ex}(a). Accordingly, the values of the dephasing rate $\Gamma_{\text{e}}$ associated with larger currents are characterized by lower occupation of the UHB, see Figs.~\ref{fig:currents_DIA_gammaes}(a) and~\ref{fig:currents_DIA_gammaes}(b). 
This result can be understood in terms of the {\em tunneling formula} for the current~\cite{mu.we.18,ma.ga.22}, namely
\begin{equation}\label{eq:tunnel_current}
 J_{\text{tun}}(\omega) = \pi t^{\ast 2} \left[ N_{\text{e}}(\omega) N_{\text{h}}(\omega+F) - N_{\text{e}}(\omega+F)N_{\text{h}}(\omega) \right],
\end{equation}
$N_{\text{h}}(\omega)\equiv A(\omega)[1-F_{\text{e}}(\omega)]$ being the hole spectral occupation function. From Eq.~\eqref{eq:tunnel_current} we see that a large occupation of the UHB causes a suppression of the current, signalling the role of $\Gamma_{\text{e}}$ as draining mechanism for the high-energy electrons which ensures the establishment of a finite steady-state current~\cite{aron.12,ma.ga.22,ma.we.23}.

\subsubsection{Spectral properties}\label{sec:results_dia_no_ph}

\begin{figure}[t]
 \includegraphics[width=\linewidth]{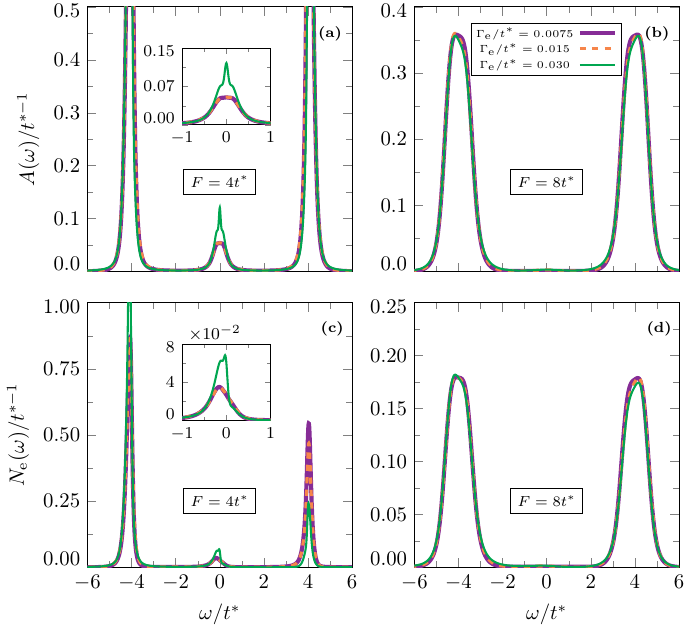}
 \caption{Electron SF $A(\omega)$ in absence of disorder at (a) $F=4t^{\ast}$ and (b) $F=8t^{\ast}$. (c) and (d) show the non-equilibrium occupation function $N_{\text{e}}(\omega)$ in Eq.~\eqref{eq:Filling_func} at the same electric field strengths. Insets in (a) and (c) show a close-up of the QPP at $\omega=0$. The values of $\Gamma_{\text{e}}$ are specified in the plot, while default parameters can be found in Table~\ref{tab:default_pars}. When present, disorder is treated in the SCB approximation. (Here $U=8t^{\ast}$ and $\overline{V^{2}}=0$.)}
 \label{fig:DIA_SFs_resonances_nodis}
\end{figure}

\begin{figure}[t]
 \includegraphics[width=\linewidth]{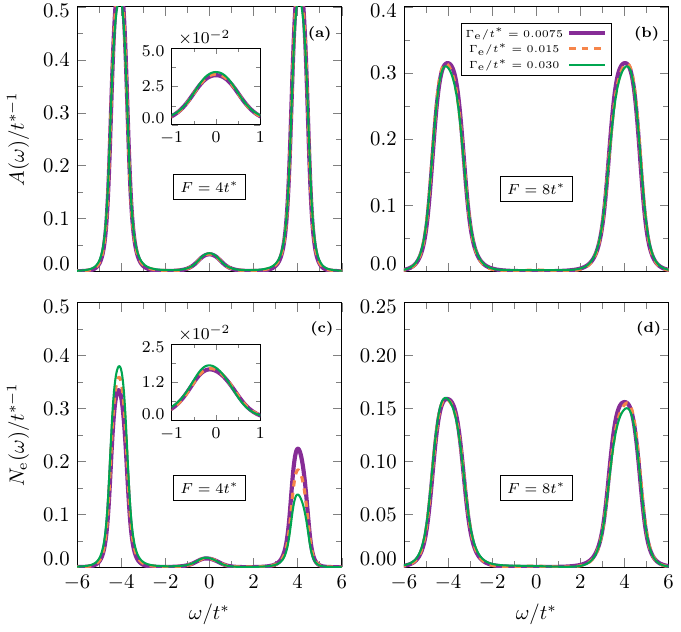}
 \caption{Same quantities as in Fig.~\ref{fig:DIA_SFs_resonances_nodis} in presence of disorder. The values of $\Gamma_{\text{e}}$ are specified in the plot, while default parameters can be found in Table~\ref{tab:default_pars}. When present, disorder is treated in the SCB approximation. (Here $U=8t^{\ast}$ and $\overline{V^{2}}=0.083t^{\ast 2}$.)}
 \label{fig:DIA_SFs_resonances}
\end{figure}

To understand the dependence of $J$ and $n_{\text{ex}}$ on the dephasing rate $\Gamma_{\text{e}}$ we focus on the spectral features at the resonances $F=U/2$ and $F=U$ both with and without disorder. In absence of disorder, at $F=U/2$ the height of the QPP at $\omega\approx 0$~\cite{aron.12,mu.we.18,ma.ga.22} is reduced when decreasing the electronic dephasing rate. 
Notably, when $\Gamma_{\text{e}}$ is smaller than $\Gamma^{\text{th}}_{\text{e}}=3\times 10^{-2}t^{\ast}$ the shape of the QPP is {\em qualitatively} different, as one can see in both $A(\omega)$ in Fig.~\ref{fig:DIA_SFs_resonances_nodis}(a) and the non-equilibrium occupation function $N_{\text{e}}(\omega)$ (Eq.~\eqref{eq:Filling_func}) shown in Fig~\ref{fig:DIA_SFs_resonances_nodis}(c). When disorder is taken into account the reduction of the QPP with a smaller $\Gamma_{\text{e}}$ can still be observed in both $A(\omega)$ and $N_{\text{e}}(\omega)$, see Figs.~\ref{fig:DIA_SFs_resonances}(a) and~\ref{fig:DIA_SFs_resonances}(c), but its overall shape is preserved and the suppression not as pronounced as in the previous case, see again Figs.~\ref{fig:DIA_SFs_resonances_nodis}(a) and~\ref{fig:DIA_SFs_resonances_nodis}(c).
However, as already pointed out in Sec.~\ref{sec:curr_charge_no_ph} (see again Figs.~\ref{fig:DIA_n_ex}(a) and~\ref{fig:DIA_n_ex}(b)) the fraction of excited electrons $n_{\text{ex}}$ around the positive-frequency region of the QPP barely changes as a function of the dephasing rate $\Gamma_{\text{e}}$ with and without disorder. In this context, the qualitative change in the QPP's spectral features in absence of disorder shown in Figs.~\ref{fig:DIA_SFs_resonances_nodis}(a) and~\ref{fig:DIA_SFs_resonances_nodis}(c) does not seem to considerably affect the conducting properties of the system.

At the resonance $F=U$ we observe that the addition of disorder does not influence the spectral features of the system appreciably, as both the SF $A(\omega)$ and the non-equilibrium occupation function $N_{\text{e}}(\omega)$ for a fixed value of the dephasing rate $\Gamma_{\text{e}}$ look alike in the two cases, see Figs.~\ref{fig:DIA_SFs_resonances_nodis}(b) and~\ref{fig:DIA_SFs_resonances}(b) for the former quantity and Figs.~\ref{fig:DIA_SFs_resonances_nodis}(d) and~\ref{fig:DIA_SFs_resonances}(d) for the latter~\footnote{We want to mention that the presence of disorder stabilizes the DMFT loop, speeding up the convergence, especially for small values of the electron dephasing rate $\Gamma_{\text{e}}$. In an interacting picture this provides more in-gap states which can help particle relaxation across the band gap, even at electric fields slightly {\em off-resonance}. For more details about the role of in-gap spectral weight in sustaining a steady-state current we refer to~\cite{aron.12,mu.we.18,ma.ga.22,ma.we.23}.}.

\subsection{Effects of e-ph interaction}

In this section we address the question of the interplay between electronic correlation and e-ph scattering mechanism in presence of disorder.

\subsubsection{Current}\label{sec:ph_and_dis_curr}

\begin{figure*}[t]
 \includegraphics[width=0.75\linewidth]{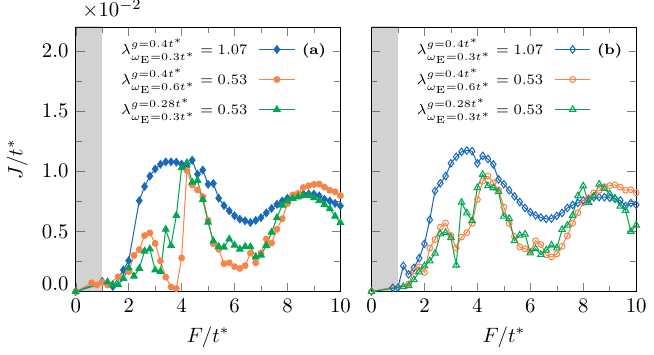}
 \caption{Current $J$ as function of the applied field $F$ with inclusion of disorder and e-ph coupling. Results are obtained for selected values of $\lambda$ (see Eq.~\eqref{eq:effective_ph_coup}) at (a) $\overline{V^{2}}=0$ and (b) $\overline{V^{2}}=0.083t^{\ast 2}$. Gray shaded rectangles denote the region showing numerical {\em instabilities}. Default parameters can be found in Table~\ref{tab:default_pars}. When present, disorder is treated in the SCB approximation. (Here $U=8t^{\ast}$ and $\Gamma_{\text{e}}=1.5\times 10^{-2}t^{\ast}$.)}
 \label{fig:DIA_currents_ph}
\end{figure*}

As in the previous setups, we start by looking at the current characteristics.

In absence of disorder and in the default parameter regime (see Table~\ref{tab:default_pars}) the $J$-$F$ curve shows the characteristic resonances at $F\approx U/3$, $U/2$ and $U$, see Fig.~\ref{fig:DIA_currents_ph}(a). Notably, enhancements in the current $J$ at off-resonant fields are observed when keeping the e-ph coupling constant $g$ at its default value and halving $\omega_{\text{\tiny E}}$. For instance, this can be detected in the merging of the two peaks at $F\approx U/3$ and $F\approx U/2$ and in the increase of $J$ for field strengths in the region $F/t^{\ast} \in \left[ 5, 8 \right]$, compare orange and blue curves in Fig.~\ref{fig:DIA_currents_ph}(a). However, a smaller $\omega_{\text{\tiny E}}$ produces, at fixed coupling $g$, a larger value of the {\em effective} e-ph coupling strength~\cite{mu.we.15,mu.ts.17}
\begin{equation}\label{eq:effective_ph_coup}
 \lambda \equiv \frac{2g^{2}}{\omega_{\text{\tiny E}}t^{\ast}}.
\end{equation}
It is, thus, interesting to assess whether this dependence on $\omega_{\text{\tiny E}}$ is intrinsic or is simply due to the corresponding change in $\lambda$. For this reason, we compare the $J$-$F$ curves in the two setups $(g,\omega_{\text{\tiny E}})$ and $(g/\sqrt{2},\omega_{\text{\tiny E}}/2)$, which are characterized by the same value of $\lambda$, see again Table~\ref{tab:default_pars} for the default parameters.

As a matter of fact, the configurations with the same $\lambda$ exhibit very similar current characteristics, as can be seen in Fig.~\ref{fig:DIA_currents_ph}(a). This suggests that the differences observed when halving $\omega_{\text{\tiny E}}$ alone are due to a stronger $\lambda$.

However, it is possible to identify the effects of the phonon frequency by direct inspection of the electronic spectral features, as we will see in Sec.~\ref{sec:ph_and_dis_sp_feat}.

In presence of disorder, the $J$-$F$ curve in the default parameters setup (see Table~\ref{tab:default_pars}) still exhibits the peaks at the main resonances $F\approx U/3$, $U/2$ and $U$, see Fig.~\ref{fig:DIA_currents_ph}(b). We also see that halving $\omega_{\text{\tiny E}}$ with respect to its default value (see Table~\ref{tab:default_pars}) still increases the current $J$ off-resonance, i.e. at $F/t^{\ast} \in \left[ 3, 4 \right]$, thus resulting in a merge of the first two resonances, and at $F/t^{\ast} \in \left[ 5, 8 \right]$, as in the case with $\overline{V^{2}}=0$. Once again, the setups characterized by the same $\lambda$ show $J$-$F$ curves which look alike, see Fig.~\ref{fig:DIA_currents_ph}(b), signalling that the current characteristics is affected by a stronger $\lambda$ also in presence of disorder.

We mention that for field strengths $F\leq t^{\ast}$, corresponding to the gray shaded rectangles in Fig.~\ref{fig:DIA_currents_ph}, the DMFT self-consistency loop is very unstable and, as such, we obtain spurious oscillations of the current $J$.
\begin{figure}[b]
 \includegraphics[width=\linewidth]{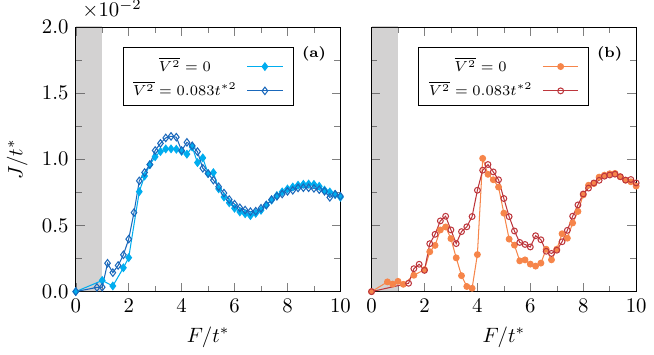}
 \caption{Current $J$ as function of the applied field for selected values of the SCB disorder amplitude $\overline{V^{2}}$ for (a) $\lambda=1.07$ ($\omega_{\text{\tiny E}}=0.3t^{\ast}$) and (b) $\lambda=0.53$ ($\omega_{\text{\tiny E}}=0.6t^{\ast}$). Gray shaded rectangles denote the region showing numerical {\em instabilities}. Default parameters can be found in Table~\ref{tab:default_pars}. When present, disorder is treated in the SCB approximation. (Here $U=8t^{\ast}$, $g=0.4t^{\ast}$ and $\Gamma_{\text{e}}=1.5\times 10^{-2}t^{\ast}$.)}
 \label{fig:DIA_currents_ph_rearranged}
\end{figure}

To conclude this section, we sort the $J$-$F$ curves by the value of $\lambda$~\footnote{We hereby recall that we vary the effective e-ph coupling $\lambda$ by changing the value of the Holstein phonon frequency $\omega_{\text{\tiny E}}$.} and compare them according to whether disorder is considered or not. As shown in Fig.~\ref{fig:DIA_currents_ph_rearranged}(a), for the larger value of $\lambda$ (corresponding to a smaller $\omega_{\text{\tiny E}}$) disorder has almost no effect on the current characteristics, except for a slight increase of $J$ at $F\approx U/2$.
On the other hand, for the smaller $\lambda$ (larger $\omega_{\text{\tiny E}}$) one notices an enhancement of $J$ when disorder is taken into account, especially in the regions $F/t^{\ast} \in \left[ 3, 4 \right]$ and $F/t^{\ast} \in \left[ 5, 7 \right]$ which were characterized by more pronounced dips in the $J$-$F$ curve for $\overline{V^{2}}=0$, see Fig.~\ref{fig:DIA_currents_ph_rearranged}(b). It is then clear that a larger value of $\lambda$ overshadows the effects due to disorder. As a matter of fact, we observe that the latter can contribute a slight increase of the current $J$ off-resonance only provided that its effects are not washed away by those of too {\em strong} phonons~\footnote{We hereby refer to the value of the effective coupling $\lambda$.}, see again Fig.~\ref{fig:DIA_currents_ph_rearranged}(b).

\subsubsection{Spectral features}\label{sec:ph_and_dis_sp_feat}

A more detailed analysis of the effect of the phonon frequency on the conducting properties requires the investigation of the electron SF and the e-ph SE. 
\begin{figure}[b]
 \includegraphics[width=\linewidth]{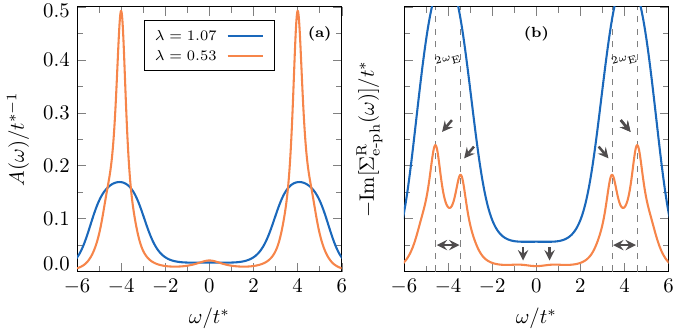}
 \caption{(a) Electron SF $A(\omega)$ and (b) imaginary part of the e-ph SE $\text{Im}\left[ \Sigma^{\text{R}}_{\text{e-ph}}(\omega)\right]$ for selected values of the Einstein phonon frequency $\omega_{\text{\tiny E}}$ at $F=3.8t^{\ast}$ and $\overline{V^{2}}=0$. Black arrows highlight the position of the subpeaks in $\text{Im}\left[ \Sigma^{\text{R}}_{\text{e-ph}}(\omega)\right]$ which are washed away by a larger $\lambda$. Default parameters can be found in Table~\ref{tab:default_pars}. When present, disorder is treated in the SCB approximation. (Here $U=8t^{\ast}$, $g=0.4t^{\ast}$ and $\Gamma_{\text{e}}=1.5\times 10^{-2}t^{\ast}$.)}
 \label{fig:DIA_spectra_ph_nodis}
\end{figure}

We stress that the latter quantity provides the main dissipation mechanisms in realistic metals, allowing electrons to get rid of their extra energy. In the following we base our analysis on a near-resonant situation, with field strengths slightly smaller or larger than half of the band gap, characterizing one of the regions where the current $J$ is enhanced by a larger {\em effective} e-ph coupling $\lambda$ both with and without disorder, as discussed in Sec.~\ref{sec:ph_and_dis_curr}. 
On the other hand, our analysis of the previous section does not display any direct dependence of the current on $\omega_{\text{\tiny E}}$ for fixed $\lambda$. By analyzing the spectral features, however, it is possible to tell the effects intrinsically related to $\omega_{\text{\tiny E}}$ from those controlled by $\lambda$, as we show below.

For $\overline{V^{2}}=0$ we observe a reduction of the height of $A(\omega)$ at $\omega \approx \pm U/2$ with consequent leak of the main Hubbard bands into the gap as $\lambda$ grows larger, see Fig.~\ref{fig:DIA_spectra_ph_nodis}(a). This, in turn, provides in-gap states which electrons can use to tunnel across the band gap even at off-resonant fields~\cite{ma.ga.22,ma.we.23}, thus enhancing the current as discussed in Sec.~\ref{sec:ph_and_dis_curr}. 

In Fig.~\ref{fig:DIA_spectra_ph_nodis}(b) we also observe that the two-peak structure in the e-ph SE $\text{Im}[ \Sigma^{R}_{\text{e-ph}}(\omega)]$ is split is completely washed away by larger values of $\lambda$. Also, a stronger e-ph coupling $\lambda$ leads to a closing of the gap in $\text{Im}[\Sigma^{R}_{\text{e-ph}}(\omega)]$ which results into more states available to electrons to relax from the upper to the lower Hubbard band, thus yielding a larger current $J$, see again the discussion in Sec.~\ref{sec:ph_and_dis_curr}. 

It is worth recalling that for electric fields slightly smaller than half of the band gap as in the situation shown in Fig.~\ref{fig:DIA_spectra_ph_nodis} the corresponding $J$-$F$ curve shows a minimum in the default parameter setup corresponding to the smaller $\lambda$ (Fig.~\ref{fig:DIA_currents_ph}(a)).

By analysing the spectral features, we see that this is due to the {\em gapped} nature of $\text{Im}[ \Sigma^{R}_{\text{e-ph}}(\omega)]$, see Fig.~\ref{fig:DIA_spectra_ph_nodis}(b), providing too few states for electrons to effectively relax to the lower Hubbard band which explains the suppression of $J$. However, despite the poor contribution of the in-gap states in the e-ph SE to the current, one can still identify the interaction between electrons and phonons in the subpeaks (separated by $2\omega_{\text{\tiny E}}$) and the small {\em satellite} peaks at $\omega\approx \pm\omega_{\text{\tiny E}}$ shown in $\text{Im}[ \Sigma^{R}_{\text{e-ph}}(\omega)]$, see Fig.~\ref{fig:DIA_spectra_ph_nodis}(b). As argued in Ref.~\cite{ma.we.23} these additional structures seem to resemble the repeated emission of phonons of frequency $\omega_{\text{\tiny E}}$ by the electrons relaxing across the band gap, see Ref.~\cite{ha.ar.23}.

On the other hand, in the setup characterized by the larger $\lambda$ the current $J$ approaches its maximum for a field strength slightly off-resonance ($F=3.8t^{\ast}$), see Fig.~\ref{fig:DIA_currents_ph}(a), which is explained by the filling of the gap observed in Fig.~\ref{fig:DIA_spectra_ph_nodis}(b). Here the finest features in $\text{Im}[\Sigma^{R}_{\text{e-ph}}(\omega)]$ observed in the previous case are now completely washed away.
\begin{figure}[b]
 \includegraphics[width=\linewidth]{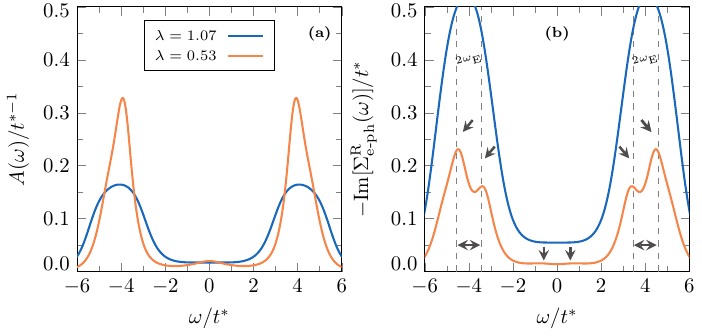}
 \caption{Same setup as in Fig~\ref{fig:DIA_spectra_ph_nodis}, in presence of disorder. Default parameters can be found in Table~\ref{tab:default_pars}. When present, disorder is treated in the SCB approximation. (Here $\overline{V^{2}}=0.083t^{\ast 2}$.)}
 \label{fig:DIA_spectra_ph_dis}
\end{figure}

When disorder is introduced ($\overline{V^{2}}=0.083t^{\ast 2}$), in the default parameter regime (see Table~\ref{tab:default_pars}) both $A(\omega)$ and $\text{Im}[\Sigma^{\text{R}}_{\text{e-ph}}(\omega)]$ are smeared out with respect to their counterparts at $\overline{V^{2}}=0$, as it can be seen by comparing Figs.~\ref{fig:DIA_spectra_ph_nodis} and~\ref{fig:DIA_spectra_ph_dis}. In addition, the spectral features corresponding to a larger $\lambda$ in Fig.~\ref{fig:DIA_spectra_ph_dis} do not differ appreciably from those in Fig.~\ref{fig:DIA_spectra_ph_nodis}, obtained in absence of disorder. 

For a finite value of the disorder amplitude $\overline{V^{2}}$, however, both $A(\omega)$ and $\text{Im}[\Sigma^{\text{R}}_{\text{e-ph}}(\omega)]$ are strongly affected by a larger value of $\lambda$. This is highlighted by the broadening of the Hubbard bands with their consequent leak into the band gap in Fig.~\ref{fig:DIA_spectra_ph_dis}(a) and the washing away of the subpeaks in the e-ph SE in Fig.~\ref{fig:DIA_spectra_ph_dis}(b), in complete analogy with the findings in absence of disorder, see Fig.~\ref{fig:DIA_spectra_ph_nodis} for comparison.

The analysis of the spectral features performed so far can explain the fundamental difference between the two dephasing channels, disorder and phonons. As a matter of fact, being introduced as an elastic scattering mechanism~\cite{ha.ja,ab.dz.75} static disorder is {\em not} expected to account for heat dissipation while phonons provide the energy dissipation mechanism in real materials. In particular, Figs.~\ref{fig:DIA_spectra_ph_nodis}(b) and~\ref{fig:DIA_spectra_ph_dis}(b) highlight the intrinsically different nature of the two mechanisms.

On the one hand, by keeping the phonon parameters fixed and including disorder we do not see any relevant changes in the in-gap states in the imaginary $\Sigma^{\text{R}}_{\text{e-ph}}$. A stronger e-ph coupling $\lambda$, instead, fills the region between the two main bands of $\text{Im}[\Sigma^{\text{R}}_{\text{e-ph}}]$ with states which are available to electrons to relax across the band gap with and without disorder. We can then argue that the elastic nature of the disorder-induced scattering mechanism manifests itself in that the regions where the spectral weight of $\text{Im}[\Sigma^{\text{R}}_{\text{e-ph}}]$ is negligible are left unaltered. Conversely, phonons contribute in-gap states so that it is easier for excited electrons to get rid of their extra energy by relaxing to the lower band. More details about this aspect will be discussed below.

\paragraph{Disorder- and phonon-induced dephasing}

This paragraph is devoted to the analysis of the dephasing effects introduced by disorder and phonons.

We start by recalling that at $\overline{V^{2}}=0$ an increased e-ph coupling $\lambda$ smears out the electron SF, see Fig.~\ref{fig:DIA_spectra_ph_nodis}(a), and washes away the finest features in the the e-ph SE $\text{Im}[\Sigma^{\text{R}}_{\text{e-ph}}(\omega)]$, see Fig.~\ref{fig:DIA_spectra_ph_nodis}(b). A similar effect (even though not as pronounced) can be observed in interacting {\em disordered} systems: below we discuss this aspect.

\begin{figure}[t]
 \includegraphics[width=\linewidth]{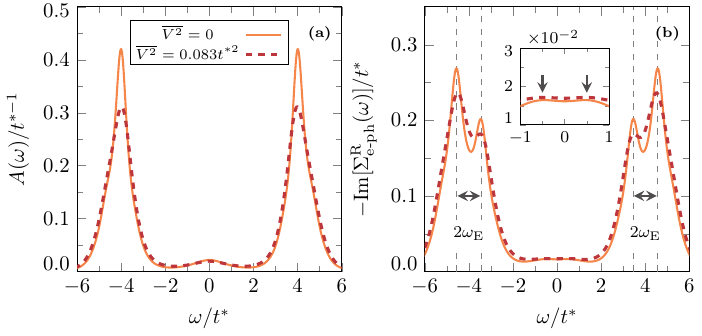}
 \caption{(a) Electron SF $A(\omega)$ and (b) corresponding imaginary part of the e-ph SE $\text{Im}[\Sigma^{\text{R}}_{\text{e-ph}}(\omega)]$ with and without disorder at $F=4.2t^{\ast}$. Black horizontal arrows denote the separation $2\omega_{\text{E}}$ between the {\em subpeaks} in the e-ph SE profile. Inset: Black vertical arrows highlight the small {\em satellite} peaks at $\omega \approx \pm \omega_{\text{E}}$. The values of $\overline{V^{2}}$ are specified in the plot, while default parameters can be found in Table~\ref{tab:default_pars}. When present, disorder is treated in the SCB approximation. (Here $U=8t^{\ast}$ and $\Gamma_{\text{e}}=1.5\times 10^{-2}t^{\ast}$.)}
 \label{fig:DIA_SFs_SEs_ph}
\end{figure}

We compare the setups in a near-resonant situation, namely $F\approx U/2$ (for fixed values of the phonon parameters) with and without disorder. In Fig.~\ref{fig:DIA_SFs_SEs_ph}(a) we display the electron SF $A(\omega)$ for the two cases. For a finite disorder amplitude ($\overline{V^{2}}=0.083t^{\ast 2}$) the main Hubbard bands are suppressed and a small fraction of their spectral weight is redistributed into the band gap. However, this effect is not as pronounced as in Fig.~\ref{fig:DIA_spectra_ph_nodis}(a), in which a stronger coupling $\lambda$ between electrons and phonons fills the band gap appreciably. Analogously, the {\em subpeaks} into which the bands of the imaginary $\Sigma^{\text{R}}_{\text{e-ph}}(\omega)$ are split are smeared out as compared to the case $\overline{V^{2}}=0$ but the small {\em satellite} peaks contributing the in-gap states are essentially left unaltered, see Fig.~\ref{fig:DIA_SFs_SEs_ph}(b). This disorder-induced broadening of the spectral features is way smaller than that shown in Fig.~\ref{fig:DIA_spectra_ph_nodis}(b) obtained with a larger $\lambda$. Once again, the findings of Fig.~\ref{fig:DIA_SFs_SEs_ph} seem to confirm that the effects of disorder can be detected mostly in the energy regions where the spectral weight is not negligible.

In analogy with the dephasing coming from the fermionic bath, see Eq.~\eqref{eq:WBL_bathGF}, we can introduce the concept of phonon and disorder dephasing rates. In particular, we define the {\em phonon dephasing rate} as
\begin{equation}\label{eq:gamma_ph-strength}
 \Gamma_{\text{ph}} = -2\text{Im}[\Sigma^{\text{R}}_{\text{e-ph}}(\omega=U/2)]|_{\tiny F=0},
\end{equation}
where the e-ph SE~\footnote{We stress that in the limit $g\to 0$ the phonon dephasing rate $\Gamma_{\text{ph}}$ is also vanishing as the e-ph SE $\Sigma^{\text{R}}_{\text{e-ph}}$ extrapolates to zero, as one can see from Eq.~(\ref{eq:backbone_e-ph_SE}).} $\Sigma^{\text{R}}_{\text{e-ph}}$ is evaluated at the characteristic energy scale provided by the UHB, i.e. $U/2$. Analogously, we introduce the {\em disorder dephasing rate} as
\begin{equation}\label{eq:gamma_dis-strength}
 \Gamma_{\text{dis}} = 2\pi \overline{V^{2}} A(\omega=U/2)|_{\tiny F=0}.
\end{equation}
The definitions in Eqs.~\eqref{eq:gamma_ph-strength} and~\eqref{eq:gamma_dis-strength} are not unique but, when computed in equilibrium ($F=0$) and at the same energy scale ($\omega\approx U/2$), provide one of the possible measures of the relative strength between the two mechanisms.

\begin{figure}[t]
 \includegraphics[width=\linewidth]{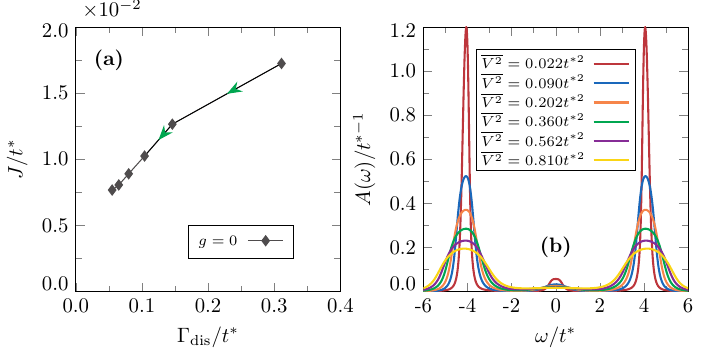}
 \caption{(a) Current $J$ as function of the {\em disorder dephasing rate} $\Gamma_{\text{dis}}$, see Eq.~\eqref{eq:gamma_dis-strength}, at $F=U/2$. Green arrows on the curve indicate the direction of increasing disorder amplitude $\overline{V^{2}}$. (b) Electron SF $A(\omega)$ corresponding to the setup in (a). Notice that increasing $\overline{V^{2}}$ (lowering $\Gamma_{\text{dis}}$) reduces the height of the Hubbard bands. Default parameters can be found in Table~\ref{tab:default_pars}. When present, disorder is treated in the SCB approximation. (Here $U=8t^{\ast}$, $\Gamma_{\text{e}}=1.5\times 10^{-2}t^{\ast}$ and $g=0$.)}
 \label{fig:scalings_static_dis_phs}
\end{figure}

It is interesting to understand whether an increased $\Gamma_{\text{dis}}$ or $\Gamma_{\text{ph}}$ can contribute an enhancement to the current in a similar way as $\Gamma_{\text{e}}$, see Sec.~\ref{sec:curr_charge_no_ph}. To do this we separately study the dependence of $J$ on $\Gamma_{\text{dis}}$ and $\Gamma_{\text{ph}}$ at $F=U/2$.

In Fig.~\ref{fig:scalings_static_dis_phs}(a) we display the current $J$ as function of the disorder dephasing rate $\Gamma_{\text{dis}}$, for a fixed $\Gamma_{\text{e}}$ and in absence of e-ph scattering. We observe an increase of the current $J$ as $\Gamma_{\text{dis}}$ grows larger~\footnote{Notice that increasing $\Gamma_{\text{dis}}$ corresponds to decreasing $\overline{V^{2}}$.}. This suggests that as far as the conducting properties are concerned the effect of an increasing disorder dephasing rate $\Gamma_{\text{dis}}$ is comparable to a larger $\Gamma_{\text{e}}$, see again Fig.~\ref{fig:currents_DIA_gammaes}(a).

This interpretation seems to be confirmed by the analysis of the SF for selected values of the SCB disorder amplitude $\overline{V^{2}}$, see Fig.~\ref{fig:scalings_static_dis_phs}(b). In fact, as for the case in Fig.~\ref{fig:DIA_SFs_resonances_nodis}(a) and especially Fig.~\ref{fig:DIA_SFs_resonances}(a), a reduced dephasing $\Gamma_{\text{dis}}$ suppresses the height of the QPP similar to what happens with a smaller $\Gamma_{\text{e}}$. In addition, an increased $\Gamma_{\text{dis}}$ suppresses the Hubbard bands, shifting some of the corresponding spectral weight near their edges, away from the QPP.

\begin{figure*}[t]
 \includegraphics[width=0.9\linewidth]{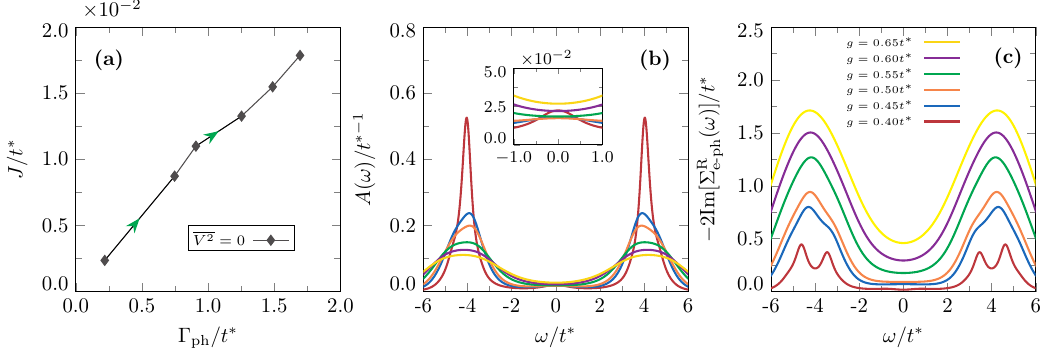}
 \caption{(a) Current $J$ as function of the {\em phonon dephasing rate} $\Gamma_{\text{ph}}$, see Eq.~\eqref{eq:gamma_ph-strength}, at $F=U/2$ in absence of disorder. Green arrows on the curve indicate the direction of increasing e-ph coupling constant $g$. (b) Electron SF $A(\omega)$ corresponding to the setup in (a). Notice that, in contrast to the case shown in Fig.~\ref{fig:scalings_static_dis_phs}, increasing $g$ corresponds to increasing $\Gamma_{\text{ph}}$, which reduces the height of the Hubbard bands. Inset shows a close-up of the in-gap states. (c) Imaginary e-ph SE $\Sigma^{\text{R}}_{\text{e-ph}}$ corresponding to (b). Default parameters can be found in Table~\ref{tab:default_pars}. When present, disorder is treated in the SCB approximation. (Here $U=8t^{\ast}$ and $\Gamma_{\text{e}}=1.5\times 10^{-2}t^{\ast}$.)}
 \label{fig:scalings_static_dis_phs2}
\end{figure*}

When disorder is neglected, at $F=U/2$ the current $J$ increases~\footnote{It should be noted that the current $J$ shows large oscillations as a function of $\Gamma_{\text{ph}}$ for $g\leq0.3t^{\ast}$.} as the phonon dephasing $\Gamma_{\text{ph}}$ grows larger~\footnote{In contrast to the case displayed in Fig.~\ref{fig:scalings_static_dis_phs}, increasing $\Gamma_{\text{ph}}$ corresponds to an increased $g$.}, for a fixed $\Gamma_{\text{e}}$, as can be seen in Fig.~\ref{fig:scalings_static_dis_phs2}(a).

The reason why the current $J$ is increased for a larger $\Gamma_{\text{ph}}$ is the creation of states around the Fermi level $\omega\approx 0$ due to the leakage of the Hubbard bands {\em directly into the gap}, as can be observed in Fig.~\ref{fig:scalings_static_dis_phs2}(b) and corresponding inset. By analyzing Fig.~\ref{fig:scalings_static_dis_phs2}(c) we can easily identify the contribution from the e-ph scattering to the dissipation as the imaginary $\Sigma^{\text{R}}_{\text{e-ph}}(\omega)$ shows a clear filling of the gap for increasing $\Gamma_{\text{ph}}$ ($g$).

\paragraph{Effect of the phonon frequency}\label{sec:ph_freq_effects}

In order to identify the effects directly related to the phonon frequency we study the changes in the spectral features when $\omega_{\text{\tiny E}}$ and $g$ are varied so to have the same value of $\lambda$.

\begin{figure*}[t]
 \includegraphics[width=0.75\linewidth]{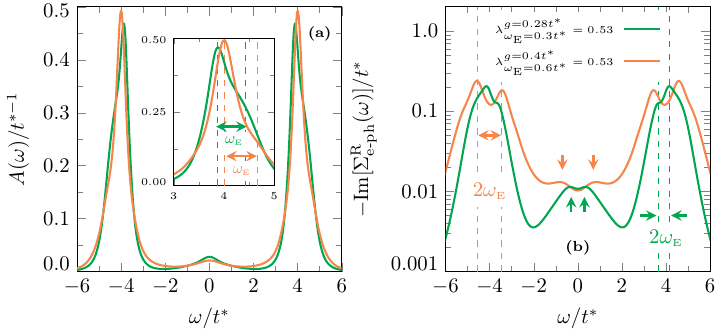}
 \caption{(a) Electron SF $A(\omega)$ for selected values of $\omega_{\text{\tiny E}}$ and $g$ at $F=3.8t^{\ast}$ and $\overline{V^{2}}=0$. Inset: Horizontal arrows highlight the distance of the small {\em shoulder} from the maximum of the UHB, roughly equal to $\omega_{\text{\tiny E}}$. (b) Imaginary e-ph SE $\text{Im}\left[ \Sigma^{\text{R}}_{\text{e-ph}}(\omega)\right]$ for the same setup. Vertical arrows point at the position of the small {\em satellite peaks} located at $\omega\approx\pm\omega_{\text{\tiny E}}$, while horizonal arrows denote the separation between the subpeaks into which each band is split, which roughly equals $2\omega_{\text{\tiny E}}$. Default parameters can be found in Table~\ref{tab:default_pars}. When present, disorder is treated in the SCB approximation. (Here $U=8t^{\ast}$, $\Gamma_{\text{e}}=1.5\times 10^{-2}t^{\ast}$ and $\lambda=0.53$.)}
 \label{fig:DIA_spectra_ph_freq_nodis}
\end{figure*}

In Fig.~\ref{fig:DIA_spectra_ph_freq_nodis} we show the differences between the configurations $(g,\omega_{\text{\tiny E}})$ and $(g/\sqrt{2},\omega_{\text{\tiny E}}/2)$, characterized by the same value of $\lambda$ at $\overline{V^{2}}=0$. Here both $A(\omega)$ and $\text{Im}[\Sigma^{\text{R}}_{\text{e-ph}}(\omega)]$ preserve their overall shape and have the same order of magnitude in the two setups. In addition, $A(\omega)$ shows a {\em tiny} shoulder which is $\sim\omega_{\text{\tiny E}}$ away from the maximum of the main Hubbard band in the two configurations, see Fig.~\ref{fig:DIA_spectra_ph_freq_nodis}(a) and inset therein. Conversely, the e-ph SE $\text{Im}[ \Sigma^{\text{R}}_{\text{e-ph}}(\omega)]$ features a splitting of its bands which amounts to $2\omega_{\text{\tiny E}}$ in both cases, see Fig.~\ref{fig:DIA_spectra_ph_freq_nodis}(b). These results do not change in presence of disorder (not shown): the spectral features are, in fact, essentially the same as in the case $\overline{V^{2}}=0$ with additional broadening coming from the disorder which makes the identification of the {\em satellite} peaks more difficult.

\section{Conclusions}\label{sec:conclusions}

In this work we characterize a {\em disordered} Mott insulating system subject to a static electric field in terms of its conducting properties.

Our main goal is to highlight the interplay between electronic correlation and e-ph interaction in the context of disordered systems, especially in the vicinity of a current-carrying regime. To do so we first focus on a purely electronic interacting system both with and witout disorder. In the first setup the current characteristics looks smoother than in the second one for all the applied field strenghts, but no substantial differences can be appreciated as for the magnitude of the current is concerned. The reason why is that disorder can only smear out the electronic spectral features by creating in-gap states near the {\em edges} of the Hubbard bands.

We then consider the influence of a self-consistent optical phonon branch interacting locally with the electrons of the lattice on top of disorder. In this setup we observe that disorder can indeed contribute a slight enhancement to the steady-state current at off-resonant applied field strengths, provided that the {\em effective} interaction among electrons and phonons is not too large. On the other hand, when the e-ph interaction is strong disorder effects cannot be appreciated.

Most importantly, we find that the two dephasing mechanisms, disorder and phonons, differ in that electron-phonon interaction provides states {\em within} the band gap while disorder only contributes a leakage of the spectral weight towards the gap which is confined to the edge on the Hubbard bands.
This crucial difference can be best appreciated when the applied field equals half of the band gap. In absence of phonons an increased disorder amplitude smears out the Hubbard bands, decreasing the height of the quasi-particle peak at around the Fermi level at the same time. This effectively suppresses the necessary states for electrons to relax to the lower band. On the other hand, when disorder is discarded, an increased electron-phonon coupling provides more states within the band gap which enhance the current by relaxing more electrons to the lower Hubbard band.

\appendix

\section{Floquet Green's function approach for static electric field}\label{sec:GFs_Dyson_Floquet}

We hereby summarize the essential aspects of the \emph{Keldysh-Floquet} formalism~\cite{ts.ok.08,sc.mo.02u,jo.fr.08}. 

In the Coulomb gauge a system subject to a constant electric field preserves the time-translational invariance~\cite{ma.ga.22,ts.ok.08}: the corresponding Green's function must obey this symmetry too. Within the temporal gauge, instead, the Green's function fulfills the periodicity condition $\kel{X}(t,t^{\prime})=\kel{X}(t+\tau,t^{\prime}+\tau)$ with $\tau=2\pi/\Omega$ being the period~\footnote{We recall that in our units $\Omega\equiv F$, see Sec.~\ref{sec:intro}.}. The Green's function can be represented as (for simplicity we drop the crystal momentum $\vec{k}$)
\begin{equation}\label{eq:FloquetGF}
\kel{X}_{mn}(\omega) =\int \dd t_{\text{rel}} \int_{-\tau/2}^{\tau/2} \frac{\dd t_{\text{av}}}{\tau} \ee^{\ii[\left(\omega+m\Omega\right) t -\left( \omega+n\Omega\right)t^{\prime}]} \kel{X}(t,t^{\prime}),
\end{equation}
where $t_{\text{rel}} = t-t^{\prime}$ and $t_{\text{av}} = (t+t^{\prime})/2$ are the relative and average time variables. Any Green's function of the form~\eqref{eq:FloquetGF} fulfills the property
\begin{equation}\label{eq:Fl_shift}
\kel{X}_{mn}(\omega) = \kel{X}_{m-n,0}(\omega+n\Omega).
\end{equation}
In Ref.~\cite{ts.ok.08} it has been shown that for a static electric field the only non-vanihing matrix elements of the {\em local} electron GF are precisely the time-averaged time-translation invariant diagonal components of Eq.~\eqref{eq:FloquetGF}. From the {\em shifting property}~\eqref{eq:Fl_shift} one can then reconstruct $\kel{X}_{nn}(\omega)$ from the knowledge of $\kel{X}_{00}(\omega)$ alone.
By performing the transformations
\begin{align}\label{eq:Fl_2_Wig_mapping}
\begin{split}
\omega^{\prime} & =\omega+(m+n)\Omega/2 \\ 
l & = m-n,
\end{split}
\end{align}
Eq.~\eqref{eq:FloquetGF} can be recast as
\begin{equation}\label{eq:WignerGF}
\kel{X}_{l}(\omega^{\prime}) =\int \dd t_{\text{rel}} \int_{-\tau/2}^{\tau/2} \frac{\dd t_{\text{av}}}{\tau} \ee^{\ii l\Omega t_{\text{av}} + \ii \omega^{\prime} t_{\text{rel}}} \kel{X}(t,t^{\prime}),
\end{equation}
which is usually referred to as the {\em Wigner representation}~\cite{ts.ok.08,ma.ga.22}.

\section{Benchmarking SCB against CPA - Part II}

\begin{figure}[t]
 \includegraphics[width=\linewidth]{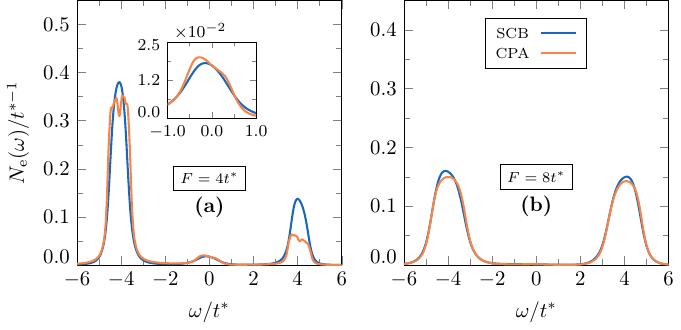}
 \caption{(a) Electron spectral occupation function $N_{\text{e}}(\omega)$ (see Eq.~\eqref{eq:Filling_func}) in the CPA and SCB schemes for $F=4t^{\ast}$. The inset shows a magnification of the region $\omega \approx 0$. (b) Same quantity for $F=8t^{\ast}$. Default parameters can be found in Table~\ref{tab:default_pars}. (Here $U=8t^{\ast}$ and $\Gamma_{\text{e}}=6\times 10^{-2}t^{\ast}$.)}
 \label{fig:DIA_CPA_app_main}
\end{figure}

In order for the comparison between the two disorder schemes performed in Sec.~\ref{sec:CPA_vs_DIA_benchmark} to be complete, in Fig.~\ref{fig:DIA_CPA_app_main} we show the electron spectral occupation function $N_{\text{e}}(\omega)$ for the two main resonances $F=U/2$, $U$. Once more, we observe that for fields compensating half of the band gap the differences in the two approaches are more pronounced, see Fig.~\ref{fig:DIA_CPA_app_main}(a), while for fully resonant fields the two curves are almost on top of each other, see Fig.~\ref{fig:DIA_CPA_app_main}(b).

\begin{figure}[b]
 \includegraphics[width=\linewidth]{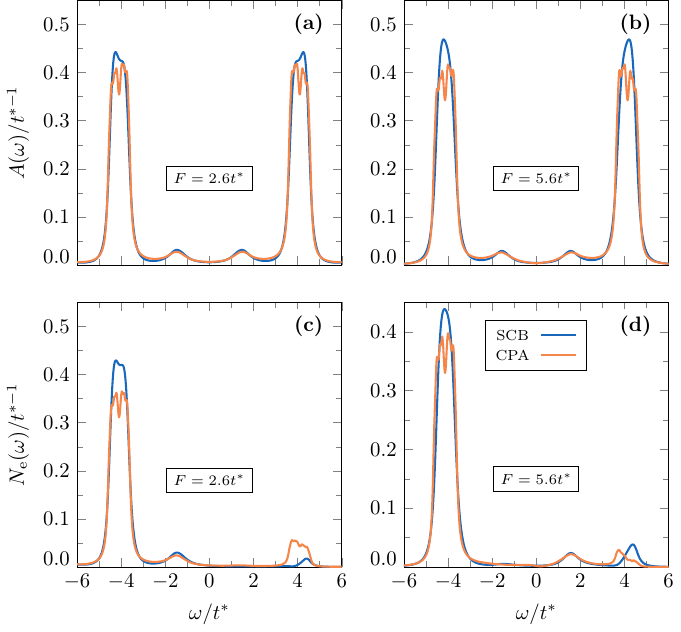}
 \caption{Electron SF in the CPA and SCB schemes for (a) $F\approx U/3$ and (b) $F\approx 2U/3$. (c) and (d) show the electron occupation function $N_{\text{e}}(\omega)$ (see Eq.~\eqref{eq:Filling_func}) for the same field strengths. Default parameters can be found in Table~\ref{tab:default_pars}. (Here $U=8t^{\ast}$ and $\Gamma_{\text{e}}=6\times 10^{-2}t^{\ast}$.)}
 \label{fig:DIA_CPA_app_side}
\end{figure}

In Fig.~\ref{fig:DIA_CPA_app_side}, instead, we show both $A(\omega)$ and $N_{\text{e}}(\omega)$ for the resonances $F\approx U/3$, $2U/3$. For both the field strengths one can observe the formation of {\em kinks} in $A(\omega)$ at the position of the Hubbard bands in the CPA scheme, see Figs.~\ref{fig:DIA_CPA_app_side}(a) and (b). The occupation function $N_{\text{e}}(\omega)$ exhibits the same differences between the two approaches, as it can be seen from Figs.~\ref{fig:DIA_CPA_app_side}(c) and (d).

\section{The non-equilibrium distribution function in absence of disorder}

Here we complement the analysis performed in Sec.~\ref{sec:results_dia_no_ph}, i.e. a setup without disorder and in absence of phonons, by showing the non-equilibrium distribution function $F_{\text{el}}(\omega)$ (see Eq.~\eqref{eq:NEFD-dist}) at the resonances $F=U/2$, $U$ for selected values of the electronic dephasing rate $\Gamma_{\text{e}}$.

\begin{figure}[t]
 \includegraphics[width=\linewidth]{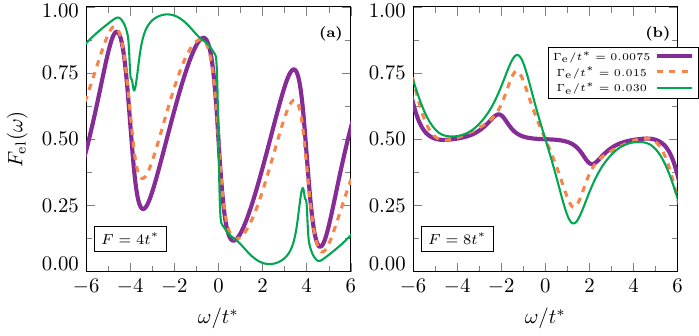}
 \caption{(a) non-equilibrium distribution function $F_{\text{el}}(\omega)$ (see Eq.~\eqref{eq:NEFD-dist}) for $F=4t^{\ast}$ at selected values of $\Gamma_{\text{e}}$. (b) Same quantity for $F=8t^{\ast}$. Default parameters can be found in Table~\ref{tab:default_pars}. (Here $U=8t^{\ast}$, $\overline{V^{2}}=0$ and $g=0$.)}
 \label{fig:DIA_app_dist}
\end{figure}

In Fig.~\ref{fig:DIA_SFs_resonances_nodis} we observed that both $A(\omega)$ and $N_{\text{e}}(\omega)$ show very sharp features for an increased electronic dephasing rate $\Gamma_{\text{e}}$, especially at $\omega \approx 0$. In Fig.~\ref{fig:DIA_app_dist} we show the corresponding non-equilibrium distribution function $F_{\text{el}}(\omega)$ for $F=U/2$ and $F=U$. In the former case, $F_{\text{el}}(\omega)$ is meaningful only in the energy regions $\omega\approx \pm U/2$ and $\omega \approx 0$, i.e. where the spectral weight is non-negligible, compare Figs.~\ref{fig:DIA_SFs_resonances_nodis}(a) and~\ref{fig:DIA_app_dist}(a). We notice that $F_{\text{el}}(\omega)$ for $\Gamma_{\text{e}}=0.03t^{\ast}$ differs appreciably from the distribution functions corresponding to lower values of $\Gamma_{\text{e}}$ shown in Fig.~\ref{fig:DIA_app_dist}(a). When the field equals the whole gap, Fig.~\ref{fig:DIA_app_dist}(b), $F_{\text{el}}(\omega)$ has not a clear interpretation, not even around the energy regions corresponding to the Hubbard bands~\footnote{In general, wherever the spectral weight vanishes the non-equilibrium distribution function $F_{\text{el}}(\omega)$ makes poor sense as it is proportioanl to the inverse of the spectrum.}, as it departs appreciably from a Fermi distribution function. In the case $F=U/2$, one can extract the effective temperatures of the Hubbard bands $T_{\text{HB}}$ and of the quasi-particle peak $T_{\text{\tiny QPP}}$ by fitting $F_{\text{el}}(\omega)$ to a Fermi function $f_{\text{\tiny F}}(\omega)= \left\{\exp \left[ \beta (\omega - \mu) \right] + 1 \right\}^{-1}$: the results are shown in Table~\ref{tab:eff_temp}.

\begin{table}[b]
  \begin{center}
\begin{tabular}{ c|cc }
      \hline
      \hline
        & $T_{\text{\tiny HB}}$ \ & $T_{\text{\tiny QPP}}$ \\
      \hline
        $\Gamma_{\text{e}} = 0.0075$ \ & $0.166$ \ & $0.141$ \\ 
        $\Gamma_{\text{e}} = 0.015$ \ & $0.201$ \ & $0.154$ \\ 
        $\Gamma_{\text{e}} = 0.030$ \ & $0.0419$ \ & $0.0439$ \\ 
      \hline
      \hline
    \end{tabular}
    \caption{Effective temperature of the Hubbard bands $T_{\text{\tiny HB}}$ and of the quasi-particle peak $T_{\text{\tiny QPP}}$ for selected values of the electronic dephasing rate $\Gamma_{\text{e}}$ at $F=U/2$. Temperatures are obtained by fitting the distribution function $F_{\text{el}}(\omega)$ in Fig.~\ref{fig:DIA_app_dist}(a) in the corresponding regions to a Fermi function. All values are given in units of $t^{\ast}$.}
    \label{tab:eff_temp}
  \end{center}
\end{table}

\begin{acknowledgments}
We thank C. Aron for contributing theoretical discussions and useful insights on how to stabilize the DMFT loop. This research was funded by the Austrian Science Fund (Grant No. P 33165-N) and by NaWi Graz. The results have been obtained using the Vienna Scientific Cluster and the D/L-Cluster Graz. \\ E.A. conceived the project and supervised the work. Code development: T.M.M. contributed disorder and phonon implementation and produced theoretical data. D.W. contributed the impurity solver. The Manuscript was drafted by T.M.M. with contributions from all authors.
\end{acknowledgments}

\bibliography{references_database,my_refs}

\begin{thebibliography}{87}%
\makeatletter
\providecommand \@ifxundefined [1]{%
 \@ifx{#1\undefined}
}%
\providecommand \@ifnum [1]{%
 \ifnum #1\expandafter \@firstoftwo
 \else \expandafter \@secondoftwo
 \fi
}%
\providecommand \@ifx [1]{%
 \ifx #1\expandafter \@firstoftwo
 \else \expandafter \@secondoftwo
 \fi
}%
\providecommand \natexlab [1]{#1}%
\providecommand \enquote  [1]{``#1''}%
\providecommand \bibnamefont  [1]{#1}%
\providecommand \bibfnamefont [1]{#1}%
\providecommand \citenamefont [1]{#1}%
\providecommand \href@noop [0]{\@secondoftwo}%
\providecommand \href [0]{\begingroup \@sanitize@url \@href}%
\providecommand \@href[1]{\@@startlink{#1}\@@href}%
\providecommand \@@href[1]{\endgroup#1\@@endlink}%
\providecommand \@sanitize@url [0]{\catcode `\\12\catcode `\$12\catcode
  `\&12\catcode `\#12\catcode `\^12\catcode `\_12\catcode `\%12\relax}%
\providecommand \@@startlink[1]{}%
\providecommand \@@endlink[0]{}%
\providecommand \url  [0]{\begingroup\@sanitize@url \@url }%
\providecommand \@url [1]{\endgroup\@href {#1}{\urlprefix }}%
\providecommand \urlprefix  [0]{URL }%
\providecommand \Eprint [0]{\href }%
\providecommand \doibase [0]{https://doi.org/}%
\providecommand \selectlanguage [0]{\@gobble}%
\providecommand \bibinfo  [0]{\@secondoftwo}%
\providecommand \bibfield  [0]{\@secondoftwo}%
\providecommand \translation [1]{[#1]}%
\providecommand \BibitemOpen [0]{}%
\providecommand \bibitemStop [0]{}%
\providecommand \bibitemNoStop [0]{.\EOS\space}%
\providecommand \EOS [0]{\spacefactor3000\relax}%
\providecommand \BibitemShut  [1]{\csname bibitem#1\endcsname}%
\let\auto@bib@innerbib\@empty
\bibitem [{\citenamefont {Stoliar}\ \emph {et~al.}(2013)\citenamefont
  {Stoliar}, \citenamefont {Cario}, \citenamefont {Janod}, \citenamefont
  {Corraze}, \citenamefont {Guillot-Deudon}, \citenamefont {Salmon-Bourmand},
  \citenamefont {Guiot}, \citenamefont {Tranchant},\ and\ \citenamefont
  {Rozenberg}}]{st.ca.13}%
  \BibitemOpen
  \bibfield  {author} {\bibinfo {author} {\bibfnamefont {P.}~\bibnamefont
  {Stoliar}}, \bibinfo {author} {\bibfnamefont {L.}~\bibnamefont {Cario}},
  \bibinfo {author} {\bibfnamefont {E.}~\bibnamefont {Janod}}, \bibinfo
  {author} {\bibfnamefont {B.}~\bibnamefont {Corraze}}, \bibinfo {author}
  {\bibfnamefont {C.}~\bibnamefont {Guillot-Deudon}}, \bibinfo {author}
  {\bibfnamefont {S.}~\bibnamefont {Salmon-Bourmand}}, \bibinfo {author}
  {\bibfnamefont {V.}~\bibnamefont {Guiot}}, \bibinfo {author} {\bibfnamefont
  {J.}~\bibnamefont {Tranchant}},\ and\ \bibinfo {author} {\bibfnamefont
  {M.}~\bibnamefont {Rozenberg}},\ }\bibfield  {title} {\bibinfo {title}
  {Universal electric-field-driven resistive transition in narrow-gap mott
  insulators},\ }\href {https://doi.org/10.1002/adma.201301113} {\bibfield
  {journal} {\bibinfo  {journal} {Advanced Materials}\ }\textbf {\bibinfo
  {volume} {25}},\ \bibinfo {pages} {3222} (\bibinfo {year}
  {2013})}\BibitemShut {NoStop}%
\bibitem [{\citenamefont {Janod}\ \emph {et~al.}(2015)\citenamefont {Janod},
  \citenamefont {Tranchant}, \citenamefont {Corraze}, \citenamefont
  {Querr{\'e}}, \citenamefont {Stoliar}, \citenamefont {Rozenberg},
  \citenamefont {Cren}, \citenamefont {Roditchev}, \citenamefont {Phuoc},
  \citenamefont {Besland},\ and\ \citenamefont {Cario}}]{ja.tr.15}%
  \BibitemOpen
  \bibfield  {author} {\bibinfo {author} {\bibfnamefont {E.}~\bibnamefont
  {Janod}}, \bibinfo {author} {\bibfnamefont {J.}~\bibnamefont {Tranchant}},
  \bibinfo {author} {\bibfnamefont {B.}~\bibnamefont {Corraze}}, \bibinfo
  {author} {\bibfnamefont {M.}~\bibnamefont {Querr{\'e}}}, \bibinfo {author}
  {\bibfnamefont {P.}~\bibnamefont {Stoliar}}, \bibinfo {author} {\bibfnamefont
  {M.}~\bibnamefont {Rozenberg}}, \bibinfo {author} {\bibfnamefont
  {T.}~\bibnamefont {Cren}}, \bibinfo {author} {\bibfnamefont {D.}~\bibnamefont
  {Roditchev}}, \bibinfo {author} {\bibfnamefont {V.~T.}\ \bibnamefont
  {Phuoc}}, \bibinfo {author} {\bibfnamefont {M.-P.}\ \bibnamefont {Besland}},\
  and\ \bibinfo {author} {\bibfnamefont {L.}~\bibnamefont {Cario}},\ }\bibfield
   {title} {\bibinfo {title} {Resistive switching in mott insulators and
  correlated systems},\ }\href {https://doi.org/10.1002/adfm.201500823}
  {\bibfield  {journal} {\bibinfo  {journal} {Advanced Functional Materials}\
  }\textbf {\bibinfo {volume} {25}},\ \bibinfo {pages} {6287} (\bibinfo {year}
  {2015})}\BibitemShut {NoStop}%
\bibitem [{\citenamefont {Lee}\ and\ \citenamefont
  {Ramakrishnan}(1985)}]{le.ra.85}%
  \BibitemOpen
  \bibfield  {author} {\bibinfo {author} {\bibfnamefont {P.~A.}\ \bibnamefont
  {Lee}}\ and\ \bibinfo {author} {\bibfnamefont {T.~V.}\ \bibnamefont
  {Ramakrishnan}},\ }\bibfield  {title} {\bibinfo {title} {Disordered
  electronic systems},\ }\href@noop {} {\bibfield  {journal} {\bibinfo
  {journal} {Rev. Mod. Phys.}\ }\textbf {\bibinfo {volume} {57}},\ \bibinfo
  {pages} {287} (\bibinfo {year} {1985})}\BibitemShut {NoStop}%
\bibitem [{\citenamefont {Belitz}\ and\ \citenamefont
  {Kirkpatrick}(1994)}]{be.ki.94}%
  \BibitemOpen
  \bibfield  {author} {\bibinfo {author} {\bibfnamefont {D.}~\bibnamefont
  {Belitz}}\ and\ \bibinfo {author} {\bibfnamefont {T.~R.}\ \bibnamefont
  {Kirkpatrick}},\ }\bibfield  {title} {\bibinfo {title} {The anderson-mott
  transition},\ }\href@noop {} {\bibfield  {journal} {\bibinfo  {journal} {Rev.
  Mod. Phys.}\ }\textbf {\bibinfo {volume} {66}},\ \bibinfo {pages} {261}
  (\bibinfo {year} {1994})}\BibitemShut {NoStop}%
\bibitem [{\citenamefont {Li}\ \emph {et~al.}(2017)\citenamefont {Li},
  \citenamefont {Aron}, \citenamefont {Kotliar},\ and\ \citenamefont
  {Han}}]{li.ar.17}%
  \BibitemOpen
  \bibfield  {author} {\bibinfo {author} {\bibfnamefont {J.}~\bibnamefont
  {Li}}, \bibinfo {author} {\bibfnamefont {C.}~\bibnamefont {Aron}}, \bibinfo
  {author} {\bibfnamefont {G.}~\bibnamefont {Kotliar}},\ and\ \bibinfo {author}
  {\bibfnamefont {J.~E.}\ \bibnamefont {Han}},\ }\bibfield  {title} {\bibinfo
  {title} {Microscopic theory of resistive switching in ordered insulators:
  Electronic versus thermal mechanisms},\ }\href
  {https://doi.org/10.1021/acs.nanolett.7b00286} {\bibfield  {journal}
  {\bibinfo  {journal} {Nano Letters}\ }\textbf {\bibinfo {volume} {17}},\
  \bibinfo {pages} {2994} (\bibinfo {year} {2017})},\ \bibinfo {note} {pMID:
  28394624}\BibitemShut {NoStop}%
\bibitem [{\citenamefont {Tsuji}\ \emph {et~al.}(2009)\citenamefont {Tsuji},
  \citenamefont {Oka},\ and\ \citenamefont {Aoki}}]{ts.ok.09}%
  \BibitemOpen
  \bibfield  {author} {\bibinfo {author} {\bibfnamefont {N.}~\bibnamefont
  {Tsuji}}, \bibinfo {author} {\bibfnamefont {T.}~\bibnamefont {Oka}},\ and\
  \bibinfo {author} {\bibfnamefont {H.}~\bibnamefont {Aoki}},\ }\bibfield
  {title} {\bibinfo {title} {Nonequilibrium steady state of photoexcited
  correlated electrons in the presence of dissipation},\ }\href
  {https://doi.org/10.1103/PhysRevLett.103.047403} {\bibfield  {journal}
  {\bibinfo  {journal} {Phys. Rev. Lett.}\ }\textbf {\bibinfo {volume} {103}},\
  \bibinfo {pages} {047403} (\bibinfo {year} {2009})}\BibitemShut {NoStop}%
\bibitem [{\citenamefont {Aron}(2012)}]{aron.12}%
  \BibitemOpen
  \bibfield  {author} {\bibinfo {author} {\bibfnamefont {C.}~\bibnamefont
  {Aron}},\ }\bibfield  {title} {\bibinfo {title} {Dielectric breakdown of a
  mott insulator},\ }\href {https://doi.org/10.1103/PhysRevB.86.085127}
  {\bibfield  {journal} {\bibinfo  {journal} {Phys. Rev. B}\ }\textbf {\bibinfo
  {volume} {86}},\ \bibinfo {pages} {085127} (\bibinfo {year}
  {2012})}\BibitemShut {NoStop}%
\bibitem [{\citenamefont {Amaricci}\ \emph {et~al.}(2012)\citenamefont
  {Amaricci}, \citenamefont {Weber}, \citenamefont {Capone},\ and\
  \citenamefont {Kotliar}}]{am.we.12}%
  \BibitemOpen
  \bibfield  {author} {\bibinfo {author} {\bibfnamefont {A.}~\bibnamefont
  {Amaricci}}, \bibinfo {author} {\bibfnamefont {C.}~\bibnamefont {Weber}},
  \bibinfo {author} {\bibfnamefont {M.}~\bibnamefont {Capone}},\ and\ \bibinfo
  {author} {\bibfnamefont {G.}~\bibnamefont {Kotliar}},\ }\bibfield  {title}
  {\bibinfo {title} {Approach to a stationary state in a driven hubbard model
  coupled to a thermostat},\ }\href
  {https://doi.org/10.1103/PhysRevB.86.085110} {\bibfield  {journal} {\bibinfo
  {journal} {Phys. Rev. B}\ }\textbf {\bibinfo {volume} {86}},\ \bibinfo
  {pages} {085110} (\bibinfo {year} {2012})}\BibitemShut {NoStop}%
\bibitem [{\citenamefont {Aron}\ \emph {et~al.}(2012)\citenamefont {Aron},
  \citenamefont {Kotliar},\ and\ \citenamefont {Weber}}]{ar.ko.12}%
  \BibitemOpen
  \bibfield  {author} {\bibinfo {author} {\bibfnamefont {C.}~\bibnamefont
  {Aron}}, \bibinfo {author} {\bibfnamefont {G.}~\bibnamefont {Kotliar}},\ and\
  \bibinfo {author} {\bibfnamefont {C.}~\bibnamefont {Weber}},\ }\bibfield
  {title} {\bibinfo {title} {Dimensional crossover driven by an electric
  field},\ }\href {https://doi.org/10.1103/PhysRevLett.108.086401} {\bibfield
  {journal} {\bibinfo  {journal} {Phys. Rev. Lett.}\ }\textbf {\bibinfo
  {volume} {108}},\ \bibinfo {pages} {086401} (\bibinfo {year}
  {2012})}\BibitemShut {NoStop}%
\bibitem [{\citenamefont {Li}\ \emph {et~al.}(2015)\citenamefont {Li},
  \citenamefont {Aron}, \citenamefont {Kotliar},\ and\ \citenamefont
  {Han}}]{li.ar.15}%
  \BibitemOpen
  \bibfield  {author} {\bibinfo {author} {\bibfnamefont {J.}~\bibnamefont
  {Li}}, \bibinfo {author} {\bibfnamefont {C.}~\bibnamefont {Aron}}, \bibinfo
  {author} {\bibfnamefont {G.}~\bibnamefont {Kotliar}},\ and\ \bibinfo {author}
  {\bibfnamefont {J.~E.}\ \bibnamefont {Han}},\ }\bibfield  {title} {\bibinfo
  {title} {Electric-field-driven resistive switching in the dissipative hubbard
  model},\ }\href {https://doi.org/10.1103/PhysRevLett.114.226403} {\bibfield
  {journal} {\bibinfo  {journal} {Phys. Rev. Lett.}\ }\textbf {\bibinfo
  {volume} {114}},\ \bibinfo {pages} {226403} (\bibinfo {year}
  {2015})}\BibitemShut {NoStop}%
\bibitem [{\citenamefont {Han}\ \emph {et~al.}(2018)\citenamefont {Han},
  \citenamefont {Li}, \citenamefont {Aron},\ and\ \citenamefont
  {Kotliar}}]{ha.li.18}%
  \BibitemOpen
  \bibfield  {author} {\bibinfo {author} {\bibfnamefont {J.~E.}\ \bibnamefont
  {Han}}, \bibinfo {author} {\bibfnamefont {J.}~\bibnamefont {Li}}, \bibinfo
  {author} {\bibfnamefont {C.}~\bibnamefont {Aron}},\ and\ \bibinfo {author}
  {\bibfnamefont {G.}~\bibnamefont {Kotliar}},\ }\bibfield  {title} {\bibinfo
  {title} {Nonequilibrium mean-field theory of resistive phase transitions},\
  }\href {https://doi.org/10.1103/PhysRevB.98.035145} {\bibfield  {journal}
  {\bibinfo  {journal} {Phys. Rev. B}\ }\textbf {\bibinfo {volume} {98}},\
  \bibinfo {pages} {035145} (\bibinfo {year} {2018})}\BibitemShut {NoStop}%
\bibitem [{\citenamefont {Murakami}\ and\ \citenamefont
  {Werner}(2018)}]{mu.we.18}%
  \BibitemOpen
  \bibfield  {author} {\bibinfo {author} {\bibfnamefont {Y.}~\bibnamefont
  {Murakami}}\ and\ \bibinfo {author} {\bibfnamefont {P.}~\bibnamefont
  {Werner}},\ }\bibfield  {title} {\bibinfo {title} {Nonequilibrium steady
  states of electric field driven mott insulators},\ }\href
  {https://doi.org/10.1103/PhysRevB.98.075102} {\bibfield  {journal} {\bibinfo
  {journal} {Phys. Rev. B}\ }\textbf {\bibinfo {volume} {98}},\ \bibinfo
  {pages} {075102} (\bibinfo {year} {2018})}\BibitemShut {NoStop}%
\bibitem [{\citenamefont {Murakami}\ \emph {et~al.}(2015)\citenamefont
  {Murakami}, \citenamefont {Werner}, \citenamefont {Tsuji},\ and\
  \citenamefont {Aoki}}]{mu.we.15}%
  \BibitemOpen
  \bibfield  {author} {\bibinfo {author} {\bibfnamefont {Y.}~\bibnamefont
  {Murakami}}, \bibinfo {author} {\bibfnamefont {P.}~\bibnamefont {Werner}},
  \bibinfo {author} {\bibfnamefont {N.}~\bibnamefont {Tsuji}},\ and\ \bibinfo
  {author} {\bibfnamefont {H.}~\bibnamefont {Aoki}},\ }\bibfield  {title}
  {\bibinfo {title} {Interaction quench in the holstein model: Thermalization
  crossover from electron- to phonon-dominated relaxation},\ }\href
  {https://doi.org/10.1103/PhysRevB.91.045128} {\bibfield  {journal} {\bibinfo
  {journal} {Phys. Rev. B}\ }\textbf {\bibinfo {volume} {91}},\ \bibinfo
  {pages} {045128} (\bibinfo {year} {2015})}\BibitemShut {NoStop}%
\bibitem [{\citenamefont {Murakami}\ \emph {et~al.}(2017)\citenamefont
  {Murakami}, \citenamefont {Tsuji}, \citenamefont {Eckstein},\ and\
  \citenamefont {Werner}}]{mu.ts.17}%
  \BibitemOpen
  \bibfield  {author} {\bibinfo {author} {\bibfnamefont {Y.}~\bibnamefont
  {Murakami}}, \bibinfo {author} {\bibfnamefont {N.}~\bibnamefont {Tsuji}},
  \bibinfo {author} {\bibfnamefont {M.}~\bibnamefont {Eckstein}},\ and\
  \bibinfo {author} {\bibfnamefont {P.}~\bibnamefont {Werner}},\ }\bibfield
  {title} {\bibinfo {title} {Nonequilibrium steady states and transient
  dynamics of conventional superconductors under phonon driving},\ }\href
  {https://doi.org/10.1103/PhysRevB.96.045125} {\bibfield  {journal} {\bibinfo
  {journal} {Phys. Rev. B}\ }\textbf {\bibinfo {volume} {96}},\ \bibinfo
  {pages} {045125} (\bibinfo {year} {2017})}\BibitemShut {NoStop}%
\bibitem [{\citenamefont {Picano}\ \emph
  {et~al.}(2023{\natexlab{a}})\citenamefont {Picano}, \citenamefont {Grandi},\
  and\ \citenamefont {Eckstein}}]{pi.gr.23}%
  \BibitemOpen
  \bibfield  {author} {\bibinfo {author} {\bibfnamefont {A.}~\bibnamefont
  {Picano}}, \bibinfo {author} {\bibfnamefont {F.}~\bibnamefont {Grandi}},\
  and\ \bibinfo {author} {\bibfnamefont {M.}~\bibnamefont {Eckstein}},\
  }\bibfield  {title} {\bibinfo {title} {Inhomogeneous disordering at a
  photoinduced charge density wave transition},\ }\href
  {https://doi.org/10.1103/PhysRevB.107.245112} {\bibfield  {journal} {\bibinfo
   {journal} {Phys. Rev. B}\ }\textbf {\bibinfo {volume} {107}},\ \bibinfo
  {pages} {245112} (\bibinfo {year} {2023}{\natexlab{a}})}\BibitemShut
  {NoStop}%
\bibitem [{\citenamefont {Picano}\ \emph
  {et~al.}(2023{\natexlab{b}})\citenamefont {Picano}, \citenamefont {Grandi},
  \citenamefont {Werner},\ and\ \citenamefont {Eckstein}}]{pi.gr.23.ss}%
  \BibitemOpen
  \bibfield  {author} {\bibinfo {author} {\bibfnamefont {A.}~\bibnamefont
  {Picano}}, \bibinfo {author} {\bibfnamefont {F.}~\bibnamefont {Grandi}},
  \bibinfo {author} {\bibfnamefont {P.}~\bibnamefont {Werner}},\ and\ \bibinfo
  {author} {\bibfnamefont {M.}~\bibnamefont {Eckstein}},\ }\bibfield  {title}
  {\bibinfo {title} {Stochastic semiclassical theory for nonequilibrium
  electron-phonon coupled systems},\ }\href
  {https://doi.org/10.1103/PhysRevB.108.035115} {\bibfield  {journal} {\bibinfo
   {journal} {Phys. Rev. B}\ }\textbf {\bibinfo {volume} {108}},\ \bibinfo
  {pages} {035115} (\bibinfo {year} {2023}{\natexlab{b}})}\BibitemShut
  {NoStop}%
\bibitem [{\citenamefont {Mazzocchi}\ \emph {et~al.}(2022)\citenamefont
  {Mazzocchi}, \citenamefont {Gazzaneo}, \citenamefont {Lotze},\ and\
  \citenamefont {Arrigoni}}]{ma.ga.22}%
  \BibitemOpen
  \bibfield  {author} {\bibinfo {author} {\bibfnamefont {T.~M.}\ \bibnamefont
  {Mazzocchi}}, \bibinfo {author} {\bibfnamefont {P.}~\bibnamefont {Gazzaneo}},
  \bibinfo {author} {\bibfnamefont {J.}~\bibnamefont {Lotze}},\ and\ \bibinfo
  {author} {\bibfnamefont {E.}~\bibnamefont {Arrigoni}},\ }\bibfield  {title}
  {\bibinfo {title} {Correlated mott insulators in strong electric fields: Role
  of phonons in heat dissipation},\ }\href
  {https://doi.org/10.1103/PhysRevB.106.125123} {\bibfield  {journal} {\bibinfo
   {journal} {Phys. Rev. B}\ }\textbf {\bibinfo {volume} {106}},\ \bibinfo
  {pages} {125123} (\bibinfo {year} {2022})}\BibitemShut {NoStop}%
\bibitem [{\citenamefont {Gazzaneo}\ \emph {et~al.}(2022)\citenamefont
  {Gazzaneo}, \citenamefont {Mazzocchi}, \citenamefont {Lotze},\ and\
  \citenamefont {Arrigoni}}]{ga.ma.22}%
  \BibitemOpen
  \bibfield  {author} {\bibinfo {author} {\bibfnamefont {P.}~\bibnamefont
  {Gazzaneo}}, \bibinfo {author} {\bibfnamefont {T.~M.}\ \bibnamefont
  {Mazzocchi}}, \bibinfo {author} {\bibfnamefont {J.}~\bibnamefont {Lotze}},\
  and\ \bibinfo {author} {\bibfnamefont {E.}~\bibnamefont {Arrigoni}},\
  }\bibfield  {title} {\bibinfo {title} {Impact ionization processes in a
  photodriven mott insulator: Influence of phononic dissipation},\ }\href
  {https://doi.org/10.1103/PhysRevB.106.195140} {\bibfield  {journal} {\bibinfo
   {journal} {Phys. Rev. B}\ }\textbf {\bibinfo {volume} {106}},\ \bibinfo
  {pages} {195140} (\bibinfo {year} {2022})}\BibitemShut {NoStop}%
\bibitem [{\citenamefont {Mazzocchi}\ \emph {et~al.}(2023)\citenamefont
  {Mazzocchi}, \citenamefont {Werner}, \citenamefont {Gazzaneo},\ and\
  \citenamefont {Arrigoni}}]{ma.we.23}%
  \BibitemOpen
  \bibfield  {author} {\bibinfo {author} {\bibfnamefont {T.~M.}\ \bibnamefont
  {Mazzocchi}}, \bibinfo {author} {\bibfnamefont {D.}~\bibnamefont {Werner}},
  \bibinfo {author} {\bibfnamefont {P.}~\bibnamefont {Gazzaneo}},\ and\
  \bibinfo {author} {\bibfnamefont {E.}~\bibnamefont {Arrigoni}},\ }\bibfield
  {title} {\bibinfo {title} {Correlated mott insulators in a strong electric
  field: The effects of phonon renormalization},\ }\href
  {https://doi.org/10.1103/PhysRevB.107.155103} {\bibfield  {journal} {\bibinfo
   {journal} {Phys. Rev. B}\ }\textbf {\bibinfo {volume} {107}},\ \bibinfo
  {pages} {155103} (\bibinfo {year} {2023})}\BibitemShut {NoStop}%
\bibitem [{\citenamefont {Han}\ \emph {et~al.}(2023)\citenamefont {Han},
  \citenamefont {Aron}, \citenamefont {Chen}, \citenamefont {Mansaray},
  \citenamefont {Han}, \citenamefont {Kim}, \citenamefont {Randle},\ and\
  \citenamefont {Bird}}]{ha.ar.23}%
  \BibitemOpen
  \bibfield  {author} {\bibinfo {author} {\bibfnamefont {J.~E.}\ \bibnamefont
  {Han}}, \bibinfo {author} {\bibfnamefont {C.}~\bibnamefont {Aron}}, \bibinfo
  {author} {\bibfnamefont {X.}~\bibnamefont {Chen}}, \bibinfo {author}
  {\bibfnamefont {I.}~\bibnamefont {Mansaray}}, \bibinfo {author}
  {\bibfnamefont {J.-H.}\ \bibnamefont {Han}}, \bibinfo {author} {\bibfnamefont
  {K.-S.}\ \bibnamefont {Kim}}, \bibinfo {author} {\bibfnamefont
  {M.}~\bibnamefont {Randle}},\ and\ \bibinfo {author} {\bibfnamefont {J.~P.}\
  \bibnamefont {Bird}},\ }\bibfield  {title} {\bibinfo {title} {Correlated
  insulator collapse due to quantum avalanche via in-gap ladder states},\
  }\bibfield  {journal} {\bibinfo  {journal} {Nature Communications}\ }\textbf
  {\bibinfo {volume} {14}},\ \href {https://doi.org/10.1038/s41467-023-38557-8}
  {10.1038/s41467-023-38557-8} (\bibinfo {year} {2023})\BibitemShut {NoStop}%
\bibitem [{\citenamefont {D\'{\i}az}\ \emph {et~al.}(2023)\citenamefont
  {D\'{\i}az}, \citenamefont {Han},\ and\ \citenamefont {Aron}}]{di.ha.23}%
  \BibitemOpen
  \bibfield  {author} {\bibinfo {author} {\bibfnamefont {M.~I.}\ \bibnamefont
  {D\'{\i}az}}, \bibinfo {author} {\bibfnamefont {J.~E.}\ \bibnamefont {Han}},\
  and\ \bibinfo {author} {\bibfnamefont {C.}~\bibnamefont {Aron}},\ }\bibfield
  {title} {\bibinfo {title} {Electrically driven insulator-to-metal transition
  in a correlated insulator: Electronic mechanism and thermal description},\
  }\href {https://doi.org/10.1103/PhysRevB.107.195148} {\bibfield  {journal}
  {\bibinfo  {journal} {Phys. Rev. B}\ }\textbf {\bibinfo {volume} {107}},\
  \bibinfo {pages} {195148} (\bibinfo {year} {2023})}\BibitemShut {NoStop}%
\bibitem [{\citenamefont {Metzner}\ and\ \citenamefont
  {Vollhardt}(1989)}]{me.vo.89}%
  \BibitemOpen
  \bibfield  {author} {\bibinfo {author} {\bibfnamefont {W.}~\bibnamefont
  {Metzner}}\ and\ \bibinfo {author} {\bibfnamefont {D.}~\bibnamefont
  {Vollhardt}},\ }\bibfield  {title} {\bibinfo {title} {Correlated lattice
  fermions in $d=\infty$ dimensions},\ }\href
  {https://doi.org/10.1103/PhysRevLett.62.324} {\bibfield  {journal} {\bibinfo
  {journal} {Phys. Rev. Lett.}\ }\textbf {\bibinfo {volume} {62}},\ \bibinfo
  {pages} {324} (\bibinfo {year} {1989})}\BibitemShut {NoStop}%
\bibitem [{\citenamefont {Georges}\ and\ \citenamefont
  {Kotliar}(1992)}]{ge.ko.92}%
  \BibitemOpen
  \bibfield  {author} {\bibinfo {author} {\bibfnamefont {A.}~\bibnamefont
  {Georges}}\ and\ \bibinfo {author} {\bibfnamefont {G.}~\bibnamefont
  {Kotliar}},\ }\bibfield  {title} {\bibinfo {title} {Hubbard model in infinite
  dimensions},\ }\href@noop {} {\bibfield  {journal} {\bibinfo  {journal}
  {Phys. Rev. B}\ }\textbf {\bibinfo {volume} {45}},\ \bibinfo {pages} {6479}
  (\bibinfo {year} {1992})}\BibitemShut {NoStop}%
\bibitem [{\citenamefont {Georges}\ \emph {et~al.}(1996)\citenamefont
  {Georges}, \citenamefont {Kotliar}, \citenamefont {Krauth},\ and\
  \citenamefont {Rozenberg}}]{ge.ko.96}%
  \BibitemOpen
  \bibfield  {author} {\bibinfo {author} {\bibfnamefont {A.}~\bibnamefont
  {Georges}}, \bibinfo {author} {\bibfnamefont {G.}~\bibnamefont {Kotliar}},
  \bibinfo {author} {\bibfnamefont {W.}~\bibnamefont {Krauth}},\ and\ \bibinfo
  {author} {\bibfnamefont {M.~J.}\ \bibnamefont {Rozenberg}},\ }\bibfield
  {title} {\bibinfo {title} {Dynamical mean-field theory of strongly correlated
  fermion systems and the limit of infinite dimensions},\ }\href
  {https://doi.org/10.1103/RevModPhys.68.13} {\bibfield  {journal} {\bibinfo
  {journal} {Rev. Mod. Phys.}\ }\textbf {\bibinfo {volume} {68}},\ \bibinfo
  {pages} {13} (\bibinfo {year} {1996})}\BibitemShut {NoStop}%
\bibitem [{\citenamefont {Kotliar}\ \emph {et~al.}(2006)\citenamefont
  {Kotliar}, \citenamefont {Savrasov}, \citenamefont {Haule}, \citenamefont
  {Oudovenko}, \citenamefont {Parcollet},\ and\ \citenamefont
  {Marianetti}}]{ko.sa.06}%
  \BibitemOpen
  \bibfield  {author} {\bibinfo {author} {\bibfnamefont {G.}~\bibnamefont
  {Kotliar}}, \bibinfo {author} {\bibfnamefont {S.~Y.}\ \bibnamefont
  {Savrasov}}, \bibinfo {author} {\bibfnamefont {K.}~\bibnamefont {Haule}},
  \bibinfo {author} {\bibfnamefont {V.~S.}\ \bibnamefont {Oudovenko}}, \bibinfo
  {author} {\bibfnamefont {O.}~\bibnamefont {Parcollet}},\ and\ \bibinfo
  {author} {\bibfnamefont {C.~A.}\ \bibnamefont {Marianetti}},\ }\bibfield
  {title} {\bibinfo {title} {Electronic structure calculations with dynamical
  mean-field theory},\ }\href {https://doi.org/10.1103/RevModPhys.78.865}
  {\bibfield  {journal} {\bibinfo  {journal} {Rev. Mod. Phys.}\ }\textbf
  {\bibinfo {volume} {78}},\ \bibinfo {pages} {865} (\bibinfo {year}
  {2006})}\BibitemShut {NoStop}%
\bibitem [{\citenamefont {Freericks}\ \emph {et~al.}(2006)\citenamefont
  {Freericks}, \citenamefont {Turkowski},\ and\ \citenamefont
  {Zlati{\'{c}}}}]{fr.tu.06}%
  \BibitemOpen
  \bibfield  {author} {\bibinfo {author} {\bibfnamefont {J.~K.}\ \bibnamefont
  {Freericks}}, \bibinfo {author} {\bibfnamefont {V.~M.}\ \bibnamefont
  {Turkowski}},\ and\ \bibinfo {author} {\bibfnamefont {V.}~\bibnamefont
  {Zlati{\'{c}}}},\ }\bibfield  {title} {\bibinfo {title} {Nonequilibrium
  dynamical mean-field theory},\ }\href
  {https://doi.org/10.1103/PhysRevLett.97.266408} {\bibfield  {journal}
  {\bibinfo  {journal} {Phys. Rev. Lett.}\ }\textbf {\bibinfo {volume} {97}},\
  \bibinfo {pages} {266408} (\bibinfo {year} {2006})}\BibitemShut {NoStop}%
\bibitem [{\citenamefont {Aoki}\ \emph {et~al.}(2014)\citenamefont {Aoki},
  \citenamefont {Tsuji}, \citenamefont {Eckstein}, \citenamefont {Kollar},
  \citenamefont {Oka},\ and\ \citenamefont {Werner}}]{ao.ts.14}%
  \BibitemOpen
  \bibfield  {author} {\bibinfo {author} {\bibfnamefont {H.}~\bibnamefont
  {Aoki}}, \bibinfo {author} {\bibfnamefont {N.}~\bibnamefont {Tsuji}},
  \bibinfo {author} {\bibfnamefont {M.}~\bibnamefont {Eckstein}}, \bibinfo
  {author} {\bibfnamefont {M.}~\bibnamefont {Kollar}}, \bibinfo {author}
  {\bibfnamefont {T.}~\bibnamefont {Oka}},\ and\ \bibinfo {author}
  {\bibfnamefont {P.}~\bibnamefont {Werner}},\ }\bibfield  {title} {\bibinfo
  {title} {Nonequilibrium dynamical mean-field theory and its applications},\
  }\href {https://doi.org/10.1103/RevModPhys.86.779} {\bibfield  {journal}
  {\bibinfo  {journal} {Rev. Mod. Phys.}\ }\textbf {\bibinfo {volume} {86}},\
  \bibinfo {pages} {779} (\bibinfo {year} {2014})}\BibitemShut {NoStop}%
\bibitem [{\citenamefont {Zunger}\ \emph {et~al.}(1990)\citenamefont {Zunger},
  \citenamefont {Wei}, \citenamefont {Ferreira},\ and\ \citenamefont
  {Bernard}}]{zu.we.90}%
  \BibitemOpen
  \bibfield  {author} {\bibinfo {author} {\bibfnamefont {A.}~\bibnamefont
  {Zunger}}, \bibinfo {author} {\bibfnamefont {S.-H.}\ \bibnamefont {Wei}},
  \bibinfo {author} {\bibfnamefont {L.~G.}\ \bibnamefont {Ferreira}},\ and\
  \bibinfo {author} {\bibfnamefont {J.~E.}\ \bibnamefont {Bernard}},\
  }\bibfield  {title} {\bibinfo {title} {Special quasirandom structures},\
  }\href {https://doi.org/10.1103/PhysRevLett.65.353} {\bibfield  {journal}
  {\bibinfo  {journal} {Phys. Rev. Lett.}\ }\textbf {\bibinfo {volume} {65}},\
  \bibinfo {pages} {353} (\bibinfo {year} {1990})}\BibitemShut {NoStop}%
\bibitem [{\citenamefont {Gonis}(1992)}]{goni.92}%
  \BibitemOpen
  \bibfield  {author} {\bibinfo {author} {\bibfnamefont {A.}~\bibnamefont
  {Gonis}},\ }\href {https://books.google.at/books?id=SqrvAAAAMAAJ} {\emph
  {\bibinfo {title} {Green Functions for Ordered and Disordered Systems}}},\
  Studies in mathematical physics\ (\bibinfo  {publisher} {North-Holland},\
  \bibinfo {year} {1992})\BibitemShut {NoStop}%
\bibitem [{\citenamefont {Anderson}(1958)}]{ande.58}%
  \BibitemOpen
  \bibfield  {author} {\bibinfo {author} {\bibfnamefont {P.~W.}\ \bibnamefont
  {Anderson}},\ }\bibfield  {title} {\bibinfo {title} {Absence of diffusion in
  certain random lattices},\ }\href {https://doi.org/10.1103/PhysRev.109.1492}
  {\bibfield  {journal} {\bibinfo  {journal} {Phys. Rev.}\ }\textbf {\bibinfo
  {volume} {109}},\ \bibinfo {pages} {1492} (\bibinfo {year}
  {1958})}\BibitemShut {NoStop}%
\bibitem [{\citenamefont {Mott}(1949)}]{mott.49}%
  \BibitemOpen
  \bibfield  {author} {\bibinfo {author} {\bibfnamefont {N.~F.}\ \bibnamefont
  {Mott}},\ }\bibfield  {title} {\bibinfo {title} {The basis of the electron
  theory of metals, with special reference to the transition metals},\ }\href
  {https://doi.org/10.1088/0370-1298/62/7/303} {\bibfield  {journal} {\bibinfo
  {journal} {Proceedings of the Physical Society. Section A}\ }\textbf
  {\bibinfo {volume} {62}},\ \bibinfo {pages} {416} (\bibinfo {year}
  {1949})}\BibitemShut {NoStop}%
\bibitem [{\citenamefont {Imada}\ \emph {et~al.}(1998)\citenamefont {Imada},
  \citenamefont {Fujimori},\ and\ \citenamefont {Tokura}}]{im.fu.98}%
  \BibitemOpen
  \bibfield  {author} {\bibinfo {author} {\bibfnamefont {M.}~\bibnamefont
  {Imada}}, \bibinfo {author} {\bibfnamefont {A.}~\bibnamefont {Fujimori}},\
  and\ \bibinfo {author} {\bibfnamefont {Y.}~\bibnamefont {Tokura}},\
  }\bibfield  {title} {\bibinfo {title} {Metal-insulator transitions and
  correlated metals in d-electron systems},\ }\href
  {https://doi.org/10.1103/RevModPhys.70.1039} {\bibfield  {journal} {\bibinfo
  {journal} {Rev. Mod. Phys.}\ }\textbf {\bibinfo {volume} {70}},\ \bibinfo
  {pages} {1039} (\bibinfo {year} {1998})}\BibitemShut {NoStop}%
\bibitem [{\citenamefont {Evers}\ and\ \citenamefont
  {Mirlin}(2008)}]{ev.me.08}%
  \BibitemOpen
  \bibfield  {author} {\bibinfo {author} {\bibfnamefont {F.}~\bibnamefont
  {Evers}}\ and\ \bibinfo {author} {\bibfnamefont {A.~D.}\ \bibnamefont
  {Mirlin}},\ }\bibfield  {title} {\bibinfo {title} {Anderson transitions},\
  }\href {https://doi.org/10.1103/RevModPhys.80.1355} {\bibfield  {journal}
  {\bibinfo  {journal} {Rev. Mod. Phys.}\ }\textbf {\bibinfo {volume} {80}},\
  \bibinfo {pages} {1355} (\bibinfo {year} {2008})}\BibitemShut {NoStop}%
\bibitem [{\citenamefont {Terletska}\ \emph {et~al.}(2018)\citenamefont
  {Terletska}, \citenamefont {Zhang}, \citenamefont {Tam}, \citenamefont
  {Berlijn}, \citenamefont {Chioncel}, \citenamefont {Vidhyadhiraja},\ and\
  \citenamefont {Jarrell}}]{te.zh.18}%
  \BibitemOpen
  \bibfield  {author} {\bibinfo {author} {\bibfnamefont {H.}~\bibnamefont
  {Terletska}}, \bibinfo {author} {\bibfnamefont {Y.}~\bibnamefont {Zhang}},
  \bibinfo {author} {\bibfnamefont {K.-M.}\ \bibnamefont {Tam}}, \bibinfo
  {author} {\bibfnamefont {T.}~\bibnamefont {Berlijn}}, \bibinfo {author}
  {\bibfnamefont {L.}~\bibnamefont {Chioncel}}, \bibinfo {author}
  {\bibfnamefont {N.}~\bibnamefont {Vidhyadhiraja}},\ and\ \bibinfo {author}
  {\bibfnamefont {M.}~\bibnamefont {Jarrell}},\ }\bibfield  {title} {\bibinfo
  {title} {Systematic quantum cluster typical medium method for the study of
  localization in strongly disordered electronic systems},\ }\href
  {https://doi.org/10.3390/app8122401} {\bibfield  {journal} {\bibinfo
  {journal} {Appl. Sci.}\ }\textbf {\bibinfo {volume} {8}},\ \bibinfo {pages}
  {2401} (\bibinfo {year} {2018})}\BibitemShut {NoStop}%
\bibitem [{\citenamefont {Basko}\ \emph {et~al.}(2006)\citenamefont {Basko},
  \citenamefont {Aleiner},\ and\ \citenamefont {Altshuler}}]{ba.al.06}%
  \BibitemOpen
  \bibfield  {author} {\bibinfo {author} {\bibfnamefont {D.}~\bibnamefont
  {Basko}}, \bibinfo {author} {\bibfnamefont {I.}~\bibnamefont {Aleiner}},\
  and\ \bibinfo {author} {\bibfnamefont {B.}~\bibnamefont {Altshuler}},\
  }\bibfield  {title} {\bibinfo {title} {Metal–insulator transition in a
  weakly interacting many-electron system with localized single-particle
  states},\ }\href {https://doi.org/https://doi.org/10.1016/j.aop.2005.11.014}
  {\bibfield  {journal} {\bibinfo  {journal} {Annals of Physics}\ }\textbf
  {\bibinfo {volume} {321}},\ \bibinfo {pages} {1126} (\bibinfo {year}
  {2006})}\BibitemShut {NoStop}%
\bibitem [{\citenamefont {Oganesyan}\ and\ \citenamefont
  {Huse}(2007)}]{og.hu.07}%
  \BibitemOpen
  \bibfield  {author} {\bibinfo {author} {\bibfnamefont {V.}~\bibnamefont
  {Oganesyan}}\ and\ \bibinfo {author} {\bibfnamefont {D.~A.}\ \bibnamefont
  {Huse}},\ }\bibfield  {title} {\bibinfo {title} {Localization of interacting
  fermions at high temperature},\ }\href
  {https://doi.org/10.1103/PhysRevB.75.155111} {\bibfield  {journal} {\bibinfo
  {journal} {Phys. Rev. B}\ }\textbf {\bibinfo {volume} {75}},\ \bibinfo
  {pages} {155111} (\bibinfo {year} {2007})}\BibitemShut {NoStop}%
\bibitem [{\citenamefont {Soven}(1967)}]{sove.67}%
  \BibitemOpen
  \bibfield  {author} {\bibinfo {author} {\bibfnamefont {P.}~\bibnamefont
  {Soven}},\ }\bibfield  {title} {\bibinfo {title} {Coherent-potential model of
  substitutional disordered alloys},\ }\href
  {https://doi.org/10.1103/PhysRev.156.809} {\bibfield  {journal} {\bibinfo
  {journal} {Phys. Rev.}\ }\textbf {\bibinfo {volume} {156}},\ \bibinfo {pages}
  {809} (\bibinfo {year} {1967})}\BibitemShut {NoStop}%
\bibitem [{\citenamefont {Velick\'y}\ \emph {et~al.}(1968)\citenamefont
  {Velick\'y}, \citenamefont {Kirkpatrick},\ and\ \citenamefont
  {Ehrenreich}}]{ve.ki.68}%
  \BibitemOpen
  \bibfield  {author} {\bibinfo {author} {\bibfnamefont {B.}~\bibnamefont
  {Velick\'y}}, \bibinfo {author} {\bibfnamefont {S.}~\bibnamefont
  {Kirkpatrick}},\ and\ \bibinfo {author} {\bibfnamefont {H.}~\bibnamefont
  {Ehrenreich}},\ }\bibfield  {title} {\bibinfo {title} {Single-site
  approximations in the electronic theory of simple binary alloys},\ }\href
  {https://doi.org/10.1103/PhysRev.175.747} {\bibfield  {journal} {\bibinfo
  {journal} {Phys. Rev.}\ }\textbf {\bibinfo {volume} {175}},\ \bibinfo {pages}
  {747} (\bibinfo {year} {1968})}\BibitemShut {NoStop}%
\bibitem [{\citenamefont {Elliott}\ \emph {et~al.}(1974)\citenamefont
  {Elliott}, \citenamefont {Krumhansl},\ and\ \citenamefont
  {Leath}}]{el.kr.74}%
  \BibitemOpen
  \bibfield  {author} {\bibinfo {author} {\bibfnamefont {R.~J.}\ \bibnamefont
  {Elliott}}, \bibinfo {author} {\bibfnamefont {J.~A.}\ \bibnamefont
  {Krumhansl}},\ and\ \bibinfo {author} {\bibfnamefont {P.~L.}\ \bibnamefont
  {Leath}},\ }\bibfield  {title} {\bibinfo {title} {The theory and properties
  of randomly disordered crystals and related physical systems},\ }\href
  {https://doi.org/10.1103/RevModPhys.46.465} {\bibfield  {journal} {\bibinfo
  {journal} {Rev. Mod. Phys.}\ }\textbf {\bibinfo {volume} {46}},\ \bibinfo
  {pages} {465} (\bibinfo {year} {1974})}\BibitemShut {NoStop}%
\bibitem [{\citenamefont {Jani}(1989)}]{jani.89}%
  \BibitemOpen
  \bibfield  {author} {\bibinfo {author} {\bibfnamefont {V.}~\bibnamefont
  {Jani}},\ }\bibfield  {title} {\bibinfo {title} {Free-energy functional in
  the generalized coherent-potential approximation},\ }\href
  {https://doi.org/10.1103/PhysRevB.40.11331} {\bibfield  {journal} {\bibinfo
  {journal} {Phys. Rev. B}\ }\textbf {\bibinfo {volume} {40}},\ \bibinfo
  {pages} {11331} (\bibinfo {year} {1989})}\BibitemShut {NoStop}%
\bibitem [{\citenamefont {Janis}\ and\ \citenamefont
  {Vollhardt}(1992)}]{ja.vo.92}%
  \BibitemOpen
  \bibfield  {author} {\bibinfo {author} {\bibfnamefont {V.}~\bibnamefont
  {Janis}}\ and\ \bibinfo {author} {\bibfnamefont {D.}~\bibnamefont
  {Vollhardt}},\ }\bibfield  {title} {\bibinfo {title} {Coupling of quantum
  degrees of freedom in strongly interacting disordered electron systems},\
  }\href {https://doi.org/10.1103/PhysRevB.46.15712} {\bibfield  {journal}
  {\bibinfo  {journal} {Phys. Rev. B}\ }\textbf {\bibinfo {volume} {46}},\
  \bibinfo {pages} {15712} (\bibinfo {year} {1992})}\BibitemShut {NoStop}%
\bibitem [{\citenamefont {Dohner}\ \emph {et~al.}(2022)\citenamefont {Dohner},
  \citenamefont {Terletska}, \citenamefont {Tam}, \citenamefont {Moreno},\ and\
  \citenamefont {Fotso}}]{do.te.22}%
  \BibitemOpen
  \bibfield  {author} {\bibinfo {author} {\bibfnamefont {E.}~\bibnamefont
  {Dohner}}, \bibinfo {author} {\bibfnamefont {H.}~\bibnamefont {Terletska}},
  \bibinfo {author} {\bibfnamefont {K.-M.}\ \bibnamefont {Tam}}, \bibinfo
  {author} {\bibfnamefont {J.}~\bibnamefont {Moreno}},\ and\ \bibinfo {author}
  {\bibfnamefont {H.~F.}\ \bibnamefont {Fotso}},\ }\bibfield  {title} {\bibinfo
  {title} {Nonequilibrium $\text{DMFT}+\text{CPA}$ for correlated disordered
  systems},\ }\href {https://doi.org/10.1103/PhysRevB.106.195156} {\bibfield
  {journal} {\bibinfo  {journal} {Phys. Rev. B}\ }\textbf {\bibinfo {volume}
  {106}},\ \bibinfo {pages} {195156} (\bibinfo {year} {2022})}\BibitemShut
  {NoStop}%
\bibitem [{\citenamefont {Yan}\ and\ \citenamefont {Werner}(2023)}]{ya.we.23u}%
  \BibitemOpen
  \bibfield  {author} {\bibinfo {author} {\bibfnamefont {J.}~\bibnamefont
  {Yan}}\ and\ \bibinfo {author} {\bibfnamefont {P.}~\bibnamefont {Werner}},\
  }\bibfield  {title} {\bibinfo {title} {Dynamical mean -- field approach to
  disordered interacting systems and applications to quantum transport
  problem}} (\bibinfo {year} {2023}),\ \bibinfo {note}
  {arXiv:2306.07029}\BibitemShut {NoStop}%
\bibitem [{\citenamefont {Arrigoni}\ \emph {et~al.}(2013)\citenamefont
  {Arrigoni}, \citenamefont {Knap},\ and\ \citenamefont {von~der
  Linden}}]{ar.kn.13}%
  \BibitemOpen
  \bibfield  {author} {\bibinfo {author} {\bibfnamefont {E.}~\bibnamefont
  {Arrigoni}}, \bibinfo {author} {\bibfnamefont {M.}~\bibnamefont {Knap}},\
  and\ \bibinfo {author} {\bibfnamefont {W.}~\bibnamefont {von~der Linden}},\
  }\bibfield  {title} {\bibinfo {title} {Nonequilibrium dynamical mean field
  theory: an auxiliary quantum master equation approach},\ }\href
  {https://doi.org/10.1103/PhysRevLett.110.086403} {\bibfield  {journal}
  {\bibinfo  {journal} {Phys. Rev. Lett.}\ }\textbf {\bibinfo {volume} {110}},\
  \bibinfo {pages} {086403} (\bibinfo {year} {2013})}\BibitemShut {NoStop}%
\bibitem [{\citenamefont {Dorda}\ \emph {et~al.}(2014)\citenamefont {Dorda},
  \citenamefont {Nuss}, \citenamefont {von~der Linden},\ and\ \citenamefont
  {Arrigoni}}]{do.nu.14}%
  \BibitemOpen
  \bibfield  {author} {\bibinfo {author} {\bibfnamefont {A.}~\bibnamefont
  {Dorda}}, \bibinfo {author} {\bibfnamefont {M.}~\bibnamefont {Nuss}},
  \bibinfo {author} {\bibfnamefont {W.}~\bibnamefont {von~der Linden}},\ and\
  \bibinfo {author} {\bibfnamefont {E.}~\bibnamefont {Arrigoni}},\ }\bibfield
  {title} {\bibinfo {title} {Auxiliary master equation approach to non --
  equilibrium correlated impurities},\ }\href
  {https://doi.org/10.1103/PhysRevB.89.165105} {\bibfield  {journal} {\bibinfo
  {journal} {Phys. Rev. B}\ }\textbf {\bibinfo {volume} {89}},\ \bibinfo
  {pages} {165105} (\bibinfo {year} {2014})}\BibitemShut {NoStop}%
\bibitem [{\citenamefont {Dorda}\ \emph {et~al.}(2015)\citenamefont {Dorda},
  \citenamefont {Ganahl}, \citenamefont {Evertz}, \citenamefont {von~der
  Linden},\ and\ \citenamefont {Arrigoni}}]{do.ga.15}%
  \BibitemOpen
  \bibfield  {author} {\bibinfo {author} {\bibfnamefont {A.}~\bibnamefont
  {Dorda}}, \bibinfo {author} {\bibfnamefont {M.}~\bibnamefont {Ganahl}},
  \bibinfo {author} {\bibfnamefont {H.~G.}\ \bibnamefont {Evertz}}, \bibinfo
  {author} {\bibfnamefont {W.}~\bibnamefont {von~der Linden}},\ and\ \bibinfo
  {author} {\bibfnamefont {E.}~\bibnamefont {Arrigoni}},\ }\bibfield  {title}
  {\bibinfo {title} {Auxiliary master equation approach within matrix product
  states: Spectral properties of the nonequilibrium anderson impurity model},\
  }\href {https://doi.org/10.1103/PhysRevB.92.125145} {\bibfield  {journal}
  {\bibinfo  {journal} {Phys. Rev. B}\ }\textbf {\bibinfo {volume} {92}},\
  \bibinfo {pages} {125145} (\bibinfo {year} {2015})}\BibitemShut {NoStop}%
\bibitem [{\citenamefont {Titvinidze}\ \emph {et~al.}(2015)\citenamefont
  {Titvinidze}, \citenamefont {Dorda}, \citenamefont {von~der Linden},\ and\
  \citenamefont {Arrigoni}}]{ti.do.15}%
  \BibitemOpen
  \bibfield  {author} {\bibinfo {author} {\bibfnamefont {I.}~\bibnamefont
  {Titvinidze}}, \bibinfo {author} {\bibfnamefont {A.}~\bibnamefont {Dorda}},
  \bibinfo {author} {\bibfnamefont {W.}~\bibnamefont {von~der Linden}},\ and\
  \bibinfo {author} {\bibfnamefont {E.}~\bibnamefont {Arrigoni}},\ }\bibfield
  {title} {\bibinfo {title} {Transport through a correlated interface:
  Auxiliary master equation approach},\ }\href
  {https://doi.org/10.1103/PhysRevB.92.245125} {\bibfield  {journal} {\bibinfo
  {journal} {Phys. Rev. B}\ }\textbf {\bibinfo {volume} {92}},\ \bibinfo
  {pages} {245125} (\bibinfo {year} {2015})}\BibitemShut {NoStop}%
\bibitem [{\citenamefont {Werner}\ \emph {et~al.}(2023)\citenamefont {Werner},
  \citenamefont {Lotze},\ and\ \citenamefont {Arrigoni}}]{we.lo.23}%
  \BibitemOpen
  \bibfield  {author} {\bibinfo {author} {\bibfnamefont {D.}~\bibnamefont
  {Werner}}, \bibinfo {author} {\bibfnamefont {J.}~\bibnamefont {Lotze}},\ and\
  \bibinfo {author} {\bibfnamefont {E.}~\bibnamefont {Arrigoni}},\ }\bibfield
  {title} {\bibinfo {title} {Configuration interaction based nonequilibrium
  steady state impurity solver},\ }\href
  {https://doi.org/10.1103/PhysRevB.107.075119} {\bibfield  {journal} {\bibinfo
   {journal} {Phys. Rev. B}\ }\textbf {\bibinfo {volume} {107}},\ \bibinfo
  {pages} {075119} (\bibinfo {year} {2023})}\BibitemShut {NoStop}%
\bibitem [{\citenamefont {Joura}\ \emph {et~al.}(2008)\citenamefont {Joura},
  \citenamefont {Freericks},\ and\ \citenamefont {Pruschke}}]{jo.fr.08}%
  \BibitemOpen
  \bibfield  {author} {\bibinfo {author} {\bibfnamefont {A.~V.}\ \bibnamefont
  {Joura}}, \bibinfo {author} {\bibfnamefont {J.~K.}\ \bibnamefont
  {Freericks}},\ and\ \bibinfo {author} {\bibfnamefont {T.}~\bibnamefont
  {Pruschke}},\ }\bibfield  {title} {\bibinfo {title} {Steady-state
  nonequilibrium density of states of driven strongly correlated lattice models
  in infinite dimensions},\ }\href
  {https://doi.org/10.1103/PhysRevLett.101.196401} {\bibfield  {journal}
  {\bibinfo  {journal} {Phys. Rev. Lett.}\ }\textbf {\bibinfo {volume} {101}},\
  \bibinfo {pages} {196401} (\bibinfo {year} {2008})}\BibitemShut {NoStop}%
\bibitem [{\citenamefont {Tsuji}\ \emph {et~al.}(2008)\citenamefont {Tsuji},
  \citenamefont {Oka},\ and\ \citenamefont {Aoki}}]{ts.ok.08}%
  \BibitemOpen
  \bibfield  {author} {\bibinfo {author} {\bibfnamefont {N.}~\bibnamefont
  {Tsuji}}, \bibinfo {author} {\bibfnamefont {T.}~\bibnamefont {Oka}},\ and\
  \bibinfo {author} {\bibfnamefont {H.}~\bibnamefont {Aoki}},\ }\bibfield
  {title} {\bibinfo {title} {Correlated electron systems periodically driven
  out of equilibrium: $floquet+dmft$ formalism},\ }\href
  {https://doi.org/10.1103/PhysRevB.78.235124} {\bibfield  {journal} {\bibinfo
  {journal} {Phys. Rev. B}\ }\textbf {\bibinfo {volume} {78}},\ \bibinfo
  {pages} {235124} (\bibinfo {year} {2008})}\BibitemShut {NoStop}%
\bibitem [{\citenamefont {Sorantin}\ \emph {et~al.}(2018)\citenamefont
  {Sorantin}, \citenamefont {Dorda}, \citenamefont {Held},\ and\ \citenamefont
  {Arrigoni}}]{so.do.18}%
  \BibitemOpen
  \bibfield  {author} {\bibinfo {author} {\bibfnamefont {M.~E.}\ \bibnamefont
  {Sorantin}}, \bibinfo {author} {\bibfnamefont {A.}~\bibnamefont {Dorda}},
  \bibinfo {author} {\bibfnamefont {K.}~\bibnamefont {Held}},\ and\ \bibinfo
  {author} {\bibfnamefont {E.}~\bibnamefont {Arrigoni}},\ }\bibfield  {title}
  {\bibinfo {title} {Impact ionization processes in the steady state of a
  driven mott-insulating layer coupled to metallic leads},\ }\href
  {https://doi.org/10.1103/PhysRevB.97.115113} {\bibfield  {journal} {\bibinfo
  {journal} {Phys. Rev. B}\ }\textbf {\bibinfo {volume} {97}},\ \bibinfo
  {pages} {115113} (\bibinfo {year} {2018})}\BibitemShut {NoStop}%
\bibitem [{\citenamefont {Eckstein}\ and\ \citenamefont
  {Werner}(2013)}]{ec.we.13}%
  \BibitemOpen
  \bibfield  {author} {\bibinfo {author} {\bibfnamefont {M.}~\bibnamefont
  {Eckstein}}\ and\ \bibinfo {author} {\bibfnamefont {P.}~\bibnamefont
  {Werner}},\ }\bibfield  {title} {\bibinfo {title} {Nonequilibrium dynamical
  mean-field simulation of inhomogeneous systems},\ }\href
  {https://doi.org/10.1103/PhysRevB.88.075135} {\bibfield  {journal} {\bibinfo
  {journal} {Phys. Rev. B}\ }\textbf {\bibinfo {volume} {88}},\ \bibinfo
  {pages} {075135} (\bibinfo {year} {2013})}\BibitemShut {NoStop}%
\bibitem [{\citenamefont {Apalkov}\ and\ \citenamefont
  {Stockman}(2012)}]{ap.st.12}%
  \BibitemOpen
  \bibfield  {author} {\bibinfo {author} {\bibfnamefont {V.}~\bibnamefont
  {Apalkov}}\ and\ \bibinfo {author} {\bibfnamefont {M.~I.}\ \bibnamefont
  {Stockman}},\ }\bibfield  {title} {\bibinfo {title} {Theory of dielectric
  nanofilms in strong ultrafast optical fields},\ }\href
  {https://doi.org/10.1103/PhysRevB.86.165118} {\bibfield  {journal} {\bibinfo
  {journal} {Phys. Rev. B}\ }\textbf {\bibinfo {volume} {86}},\ \bibinfo
  {pages} {165118} (\bibinfo {year} {2012})}\BibitemShut {NoStop}%
\bibitem [{\citenamefont {Murakami}\ \emph {et~al.}(2018)\citenamefont
  {Murakami}, \citenamefont {Eckstein},\ and\ \citenamefont
  {Werner}}]{mu.ec.18}%
  \BibitemOpen
  \bibfield  {author} {\bibinfo {author} {\bibfnamefont {Y.}~\bibnamefont
  {Murakami}}, \bibinfo {author} {\bibfnamefont {M.}~\bibnamefont {Eckstein}},\
  and\ \bibinfo {author} {\bibfnamefont {P.}~\bibnamefont {Werner}},\
  }\bibfield  {title} {\bibinfo {title} {High-harmonic generation in mott
  insulators},\ }\href {https://doi.org/10.1103/PhysRevLett.121.057405}
  {\bibfield  {journal} {\bibinfo  {journal} {Phys. Rev. Lett.}\ }\textbf
  {\bibinfo {volume} {121}},\ \bibinfo {pages} {057405} (\bibinfo {year}
  {2018})}\BibitemShut {NoStop}%
\bibitem [{\citenamefont {Peierls}(1933)}]{peie.33}%
  \BibitemOpen
  \bibfield  {author} {\bibinfo {author} {\bibfnamefont {R.}~\bibnamefont
  {Peierls}},\ }\bibfield  {title} {\bibinfo {title} {Zur theorie des
  diamagnetismus von leitungselektronen},\ }\href
  {https://doi.org/10.1007/bf01342591} {\bibfield  {journal} {\bibinfo
  {journal} {Zeitschrift f\"ur Physik A Hadrons and Nuclei}\ }\textbf {\bibinfo
  {volume} {80}},\ \bibinfo {pages} {763} (\bibinfo {year} {1933})}\BibitemShut
  {NoStop}%
\bibitem [{\citenamefont {Neumayer}\ \emph {et~al.}(2015)\citenamefont
  {Neumayer}, \citenamefont {Arrigoni}, \citenamefont {Aichhorn},\ and\
  \citenamefont {von~der Linden}}]{ne.ar.15}%
  \BibitemOpen
  \bibfield  {author} {\bibinfo {author} {\bibfnamefont {J.}~\bibnamefont
  {Neumayer}}, \bibinfo {author} {\bibfnamefont {E.}~\bibnamefont {Arrigoni}},
  \bibinfo {author} {\bibfnamefont {M.}~\bibnamefont {Aichhorn}},\ and\
  \bibinfo {author} {\bibfnamefont {W.}~\bibnamefont {von~der Linden}},\
  }\bibfield  {title} {\bibinfo {title} {Current characteristics of a
  one-dimensional hubbard chain: Role of correlation and dissipation},\ }\href
  {https://doi.org/10.1103/PhysRevB.92.125149} {\bibfield  {journal} {\bibinfo
  {journal} {Phys. Rev. B}\ }\textbf {\bibinfo {volume} {92}},\ \bibinfo
  {pages} {125149} (\bibinfo {year} {2015})}\BibitemShut {NoStop}%
\bibitem [{Note1()}]{Note1}%
  \BibitemOpen
  \bibinfo {note} {The Authors are aware that this choice is quite special as
  it prevents the investigation of the so-called {\protect \em dimensional
  crossover} occurring at the IMT, see e.g. Ref.~\cite {aron.12}. A possible
  setup to investigate this effect would be a two-dimensional {\protect \em
  Bravais} lattice but that is beyond the purpose of this work.}\BibitemShut
  {Stop}%
\bibitem [{\citenamefont {Turkowski}\ and\ \citenamefont
  {Freericks}(2005)}]{tu.fr.05}%
  \BibitemOpen
  \bibfield  {author} {\bibinfo {author} {\bibfnamefont {V.}~\bibnamefont
  {Turkowski}}\ and\ \bibinfo {author} {\bibfnamefont {J.~K.}\ \bibnamefont
  {Freericks}},\ }\bibfield  {title} {\bibinfo {title} {Nonlinear response of
  bloch electrons in infinite dimensions},\ }\href
  {http://link.aps.org/abstract/PRB/v71/e085104} {\bibfield  {journal}
  {\bibinfo  {journal} {Phys. Rev. B}\ }\textbf {\bibinfo {volume} {71}},\
  \bibinfo {pages} {085104} (\bibinfo {year} {2005})}\BibitemShut {NoStop}%
\bibitem [{\citenamefont {Haug}\ and\ \citenamefont {Jauho}(1998)}]{ha.ja}%
  \BibitemOpen
  \bibfield  {author} {\bibinfo {author} {\bibfnamefont {H.}~\bibnamefont
  {Haug}}\ and\ \bibinfo {author} {\bibfnamefont {A.-P.}\ \bibnamefont
  {Jauho}},\ }\href {http://www.springer.com/us/book/9783540735618} {\emph
  {\bibinfo {title} {Quantum Kinetics in Transport and Optics of
  Semiconductors}}}\ (\bibinfo  {publisher} {Springer},\ \bibinfo {address}
  {Heidelberg},\ \bibinfo {year} {1998})\BibitemShut {NoStop}%
\bibitem [{\citenamefont {Schwinger}(1961)}]{schw.61}%
  \BibitemOpen
  \bibfield  {author} {\bibinfo {author} {\bibfnamefont {J.}~\bibnamefont
  {Schwinger}},\ }\bibfield  {title} {\bibinfo {title} {Brownian motion of a
  quantum oscillator},\ }\href@noop {} {\bibfield  {journal} {\bibinfo
  {journal} {J. Math. Phys.}\ }\textbf {\bibinfo {volume} {2}},\ \bibinfo
  {pages} {407} (\bibinfo {year} {1961})}\BibitemShut {NoStop}%
\bibitem [{\citenamefont {Keldysh}(1965)}]{keld.65}%
  \BibitemOpen
  \bibfield  {author} {\bibinfo {author} {\bibfnamefont {L.~V.}\ \bibnamefont
  {Keldysh}},\ }\bibfield  {title} {\bibinfo {title} {Diagram technique for
  nonequilibrium processes},\ }\href@noop {} {\bibfield  {journal} {\bibinfo
  {journal} {Sov. Phys. JETP}\ }\textbf {\bibinfo {volume} {20}},\ \bibinfo
  {pages} {1018} (\bibinfo {year} {1965})}\BibitemShut {NoStop}%
\bibitem [{\citenamefont {Rammer}\ and\ \citenamefont
  {Smith}(1986)}]{ra.sm.86}%
  \BibitemOpen
  \bibfield  {author} {\bibinfo {author} {\bibfnamefont {J.}~\bibnamefont
  {Rammer}}\ and\ \bibinfo {author} {\bibfnamefont {H.}~\bibnamefont {Smith}},\
  }\bibfield  {title} {\bibinfo {title} {Quantum field-theoretical methods in
  transport theory of metals},\ }\href
  {https://doi.org/10.1103/RevModPhys.58.323} {\bibfield  {journal} {\bibinfo
  {journal} {Rev. Mod. Phys.}\ }\textbf {\bibinfo {volume} {58}},\ \bibinfo
  {pages} {323} (\bibinfo {year} {1986})}\BibitemShut {NoStop}%
\bibitem [{\citenamefont {Abrikosov}\ \emph {et~al.}(1975)\citenamefont
  {Abrikosov}, \citenamefont {Dzyaloshinskii}, \citenamefont {Gorkov},\ and\
  \citenamefont {Silverman}}]{ab.dz.75}%
  \BibitemOpen
  \bibfield  {author} {\bibinfo {author} {\bibfnamefont {A.~A.}\ \bibnamefont
  {Abrikosov}}, \bibinfo {author} {\bibfnamefont {I.}~\bibnamefont
  {Dzyaloshinskii}}, \bibinfo {author} {\bibfnamefont {L.~P.}\ \bibnamefont
  {Gorkov}},\ and\ \bibinfo {author} {\bibfnamefont {R.~A.}\ \bibnamefont
  {Silverman}},\ }\href {https://cds.cern.ch/record/107441} {\emph {\bibinfo
  {title} {Methods of quantum field theory in statistical physics}}}\ (\bibinfo
   {publisher} {Dover},\ \bibinfo {address} {New York, NY},\ \bibinfo {year}
  {1975})\BibitemShut {NoStop}%
\bibitem [{Note2()}]{Note2}%
  \BibitemOpen
  \bibinfo {note} {We recall that the {\protect \em Keldysh} component in
  Eq.~(\ref {eq:WBL_bathGF}) is obtained from the {\protect \em
  fluctuation-dissipation theorem}, i.e. $\Sigma ^{\protect \text
  {K}}_{\protect \text {bath}}(\omega ) = 2\protect \ii \protect \text
  {Im}[\Sigma ^{\protect \text {R}}_{\protect \text {bath}}(\omega )] \protect
  \qopname \relax o{tanh}\left [ \beta \left (\omega -\mu \right )/2\right ]$,
  since the heat bath is at equilibrium.}\BibitemShut {Stop}%
\bibitem [{Note3()}]{Note3}%
  \BibitemOpen
  \bibinfo {note} {For details about the DMFT loop we point at Refs.~\cite
  {so.do.18,ma.ga.22,ga.ma.22,ma.we.23}.}\BibitemShut {Stop}%
\bibitem [{Note4()}]{Note4}%
  \BibitemOpen
  \bibinfo {note} {For further details about the $\protect \text {DMFT}$ loop
  we refer to~\cite {ma.ga.22} (in particular see Sec. III C therein) and to
  the flowchart in Fig.~\ref {fig:FDMFT_DIA_scheme} in this Manuscript, while
  the latest developments concerning the AMEA impurity solver have been
  discussed in Refs~\cite {we.lo.23,ma.we.23}.}\BibitemShut {Stop}%
\bibitem [{Note5()}]{Note5}%
  \BibitemOpen
  \bibinfo {note} {The size of the matrices in the Floquet sector is
  $N_{\protect \text {F}}=21$ and it has been chosen in such a way that the
  electron features are converged with respect to it.}\BibitemShut {Stop}%
\bibitem [{Note6()}]{Note6}%
  \BibitemOpen
  \bibinfo {note} {We recall that in the particle-hole symmetric case
  considered in this work $\varepsilon _{\protect \text
  {c}}=-U/2$.}\BibitemShut {Stop}%
\bibitem [{Note7()}]{Note7}%
  \BibitemOpen
  \bibinfo {note} {Notice that in the Migdal approximation the {\protect \em
  Hartree term} amounts to a constant energy shift that can be reabsorbed in
  the interacting e-ph Hamiltonian $\protect \hat {H}_{\protect \text {e-ph}}$
  at half-filling.}\BibitemShut {Stop}%
\bibitem [{Note8()}]{Note8}%
  \BibitemOpen
  \bibinfo {note} {As usual, the Keldysh component $D^{\protect \text
  {K}}_{\protect \text {ph}}(\omega )$ can be neglected due to the presence of
  the ohmic bath $\protect \underline {\Pi }_{\protect \text {ohm}}$ in
  Eq.~(\ref {eq:local_Dyson_Ein_ph}).}\BibitemShut {Stop}%
\bibitem [{Note9()}]{Note9}%
  \BibitemOpen
  \bibinfo {note} {Notice that Eq.~(\ref {eq:ohm_bath_spec}) ensures a linear
  dependence within almost the entire interval $\omega \in [-\omega _{\protect
  \text {o}},\omega _{\protect \text {o}}]$.}\BibitemShut {Stop}%
\bibitem [{Note10()}]{Note10}%
  \BibitemOpen
  \bibinfo {note} {As pointed out in Refs~\cite {ts.ok.08,ma.ga.22,ma.we.23},
  any quantity with a single index refers to the Wigner representation, which
  obeys the relation $\protect \underline {X}_{mn}(\omega )=\protect \underline
  {X}_{m-n}(\omega +(m+n)\Omega /2)$, where $X_{mn}(\omega )$ is any
  Floquet-represented matrix.}\BibitemShut {Stop}%
\bibitem [{Note11()}]{Note11}%
  \BibitemOpen
  \bibinfo {note} {To conclude we just want to mention that the region
  $F/t^{\ast } \in [0,1]$ required a higher resolution in the AMEA impurity
  solver as the {\protect \em hybridization} functions corresponding to those
  field strengths contain very fine features which, if not correctly resolved,
  could lead to artefacts in the observables. For a detailed discussion about
  the impurity solver hereby employed we refer to our recent work~\cite
  {we.lo.23,ma.we.23}.}\BibitemShut {Stop}%
\bibitem [{Note12()}]{Note12}%
  \BibitemOpen
  \bibinfo {note} {Of course, this is valid only for corresponding values of
  the parameters as identified by Eq.~(\ref
  {eq:DIA_CPA_dis_amplitudes}).}\BibitemShut {Stop}%
\bibitem [{Note13()}]{Note13}%
  \BibitemOpen
  \bibinfo {note} {This result is valid for large fields where dissipation is
  essential in order to sustain a steady state current. Below a certain
  threshold field (not shown), we expect this behavior to be
  reversed.}\BibitemShut {Stop}%
\bibitem [{Note14()}]{Note14}%
  \BibitemOpen
  \bibinfo {note} {We cannot reach down to $\Gamma _{\protect \text {e}}=0$, as
  it is hard to get to a converged solution of the DMFT loop for too small
  values of $\Gamma _{\protect \text {e}}$.}\BibitemShut {Stop}%
\bibitem [{Note15()}]{Note15}%
  \BibitemOpen
  \bibinfo {note} {At $F=U/2$ we can indeed talk about metallic phase due to
  the occurrence of the maximum of in-gap states located at $\omega =0$ which
  provide the necessary spectral weight to sustain electron tunneling across
  the band gap.}\BibitemShut {Stop}%
\bibitem [{Note16()}]{Note16}%
  \BibitemOpen
  \bibinfo {note} {We want to mention that the presence of disorder stabilizes
  the DMFT loop, speeding up the convergence, especially for small values of
  the electron dephasing rate $\Gamma _{\protect \text {e}}$. In an interacting
  picture this provides more in-gap states which can help particle relaxation
  across the band gap, even at electric fields slightly {\protect \em
  off-resonance}. For more details about the role of in-gap spectral weight in
  sustaining a steady-state current we refer to~\cite
  {aron.12,mu.we.18,ma.ga.22,ma.we.23}.}\BibitemShut {Stop}%
\bibitem [{Note17()}]{Note17}%
  \BibitemOpen
  \bibinfo {note} {We hereby recall that we vary the effective e-ph coupling
  $\lambda $ by changing the value of the Holstein phonon frequency $\omega
  _{\protect \text {\relax \protect \fontsize {5}{6}\protect \selectfont
  E}}$.}\BibitemShut {Stop}%
\bibitem [{Note18()}]{Note18}%
  \BibitemOpen
  \bibinfo {note} {We hereby refer to the value of the effective coupling
  $\lambda $.}\BibitemShut {Stop}%
\bibitem [{Note19()}]{Note19}%
  \BibitemOpen
  \bibinfo {note} {We stress that in the limit $g\to 0$ the phonon dephasing
  rate $\Gamma _{\protect \text {ph}}$ is also vanishing as the e-ph SE $\Sigma
  ^{\protect \text {R}}_{\protect \text {e-ph}}$ extrapolates to zero, as one
  can see from Eq.~(\ref {eq:backbone_e-ph_SE}).}\BibitemShut {Stop}%
\bibitem [{Note20()}]{Note20}%
  \BibitemOpen
  \bibinfo {note} {Notice that increasing $\Gamma _{\protect \text {dis}}$
  corresponds to decreasing $\protect \overline {V^{2}}$.}\BibitemShut {Stop}%
\bibitem [{Note21()}]{Note21}%
  \BibitemOpen
  \bibinfo {note} {It should be noted that the current $J$ shows large
  oscillations as a function of $\Gamma _{\protect \text {ph}}$ for $g\leq
  0.3t^{\ast }$.}\BibitemShut {Stop}%
\bibitem [{Note22()}]{Note22}%
  \BibitemOpen
  \bibinfo {note} {In contrast to the case displayed in Fig.~\ref
  {fig:scalings_static_dis_phs}, increasing $\Gamma _{\protect \text {ph}}$
  corresponds to an increased $g$.}\BibitemShut {Stop}%
\bibitem [{\citenamefont {Schmidt}\ and\ \citenamefont
  {Monien}(2002)}]{sc.mo.02u}%
  \BibitemOpen
  \bibfield  {author} {\bibinfo {author} {\bibfnamefont {P.}~\bibnamefont
  {Schmidt}}\ and\ \bibinfo {author} {\bibfnamefont {H.}~\bibnamefont
  {Monien}},\ }\bibfield  {title} {\bibinfo {title} {Nonequilibrium dynamical
  mean -- field theory of a strongly correlated system}} (\bibinfo {year}
  {2002}),\ \bibinfo {note} {cond-mat/0202046}\BibitemShut {NoStop}%
\bibitem [{Note23()}]{Note23}%
  \BibitemOpen
  \bibinfo {note} {We recall that in our units $\Omega \equiv F$, see Sec.~\ref
  {sec:intro}.}\BibitemShut {Stop}%
\bibitem [{Note24()}]{Note24}%
  \BibitemOpen
  \bibinfo {note} {In general, wherever the spectral weight vanishes the
  non-equilibrium distribution function $F_{\protect \text {el}}(\omega )$
  makes poor sense as it is proportioanl to the inverse of the
  spectrum.}\BibitemShut {Stop}%
\end{thebibliography}%

\end{document}